\documentclass[3p]{elsarticle}

\usepackage{hyperref}
\usepackage{amsmath}
\usepackage{amssymb}
\usepackage{xcolor}
\definecolor{darkblue}{rgb}{0.0, 0.0, 0.55}

\usepackage{xcolor}

\newcommand\hl[1]{%
  \bgroup
  \hskip0pt%
  #1%
  \egroup
}

\usepackage{float}
\usepackage{bm}
\usepackage{xspace}
\usepackage{rotating}
\usepackage{caption}
\usepackage{subcaption}

\usepackage[noend]{algpseudocode}

\journal{NeuroImage}

\newcommand{\mbf}[1]{\mathbf{#1}}

\newcommand{\volume}{\mathcal{V}}%
\newcommand{\voxels}{\mathbb{V}}%

\newcommand{\iodfone}{single-ODF\xspace}
\newcommand{\iodftwo}{double-ODF\xspace}

\newcommand{\beginsupplement}{%
        \setcounter{table}{0}
        \renewcommand{\thetable}{S\arabic{table}}%
        \setcounter{figure}{0}
        \renewcommand{\thefigure}{S\arabic{figure}}%
}

\renewcommand{\Pr}{\text{P}}

\biboptions{authoryear}

\begin{document}

\begin{frontmatter}

\title{Analytic tractography: A closed-form solution for estimating local white matter connectivity with diffusion MRI}

\author[add1]{Matthew Cieslak}
\ead{matthew.cieslak@psych.ucsb.edu}
\author[add2]{Tegan Brennan}
\ead{tegan@cs.ucsb.edu}
\author[add3]{Wendy Meiring}
\ead{meiring@pstat.ucsb.edu}
\author[add1,add4]{Lukas J. Volz}
\ead{volz@psych.ucsb.edu}
\author[add5]{Clint Greene}
\ead{clint@ece.ucsb.edu}
\author[add1]{Alexander Asturias}
\ead{asturias@psych.ucsb.edu}
\author[add2]{Subhash Suri}
\ead{suri@cs.ucsb.edu}
\author[add1]{Scott T. Grafton}
\ead{grafton@psych.ucsb.edu}

\address[add1]{Department of Psychological and Brain Sciences, University of California Santa Barbara}
\address[add2]{Department of Computer Science, University of California Santa Barbara}
\address[add3]{Department of Statistics and Applied Probability, University of California Santa Barbara}
\address[add4]{SAGE Center for the Study of the Mind, University of California Santa Barbara}
\address[add5]{Department of Electrical and Computer Engineering, University of California Santa Barbara}

\begin{abstract}

White matter structures composed of myelinated axons in the living
human brain are primarily studied by diffusion-weighted MRI (dMRI).
These \hl{long-range projections are typically characterized in a
two-step process: dMRI signal is used to estimate the orientation of
axon segments within each voxel, then these local orientations are
linked together to estimate the spatial extent of putative white matter
bundles. \hl{Tractography, the process of tracing bundles across
voxels, either requires computationally expensive (probabilistic)
simulations to model uncertainty in fiber orientation or ignores it
completely (deterministic).} Furthermore, simulation necessarily
generates a finite number of trajectories, introducing ``simulation
error'' to trajectory estimates.  Here we introduce a method to
analytically (via a closed-form solution) take an orientation
distribution function (ODF) from each voxel and calculate the
probabilities that a trajectory projects from a voxel into each
directly adjacent voxels.  We validate our method by demonstrating
experimentally that probabilistic simulations converge to our
analytically computed transition probabilities at the voxel level as
the number of simulated seeds increases. We then show that our method
accurately calculates the ground-truth transition probabilities from a
publicly available phantom dataset. As a demonstration, we incoroporate
our analytic method for voxel transition probabilities into the Voxel
Graph framework, creating a quantitative framework for assessing white
matter structure, which we call ``analytic tractography''. The
long-range connectivity problem is reduced to finding paths in a graph
whose adjacency structure reflects voxel-to-voxel analytic transition
probabilities.} We demonstrate that this approach performs comparably
to the current most widely-used probabilistic and deterministic
approaches at a fraction of the computational cost. We also demonstrate
that analytic tractography works on multiple diffusion sampling
schemes, reconstruction method or parameters used to define paths.
Open source software compatible with popular dMRI reconstruction
software is provided.

\end{abstract}


\end{frontmatter}

\section{Introduction}

\hl{The goal of structural connectivity research is to estimate the degree to which
regions in the brain are connected by axonal projections, typically by making
inferences on diffusion weighted MRI (dMRI) data. Any MRI-based method
captures signal within a discrete set of voxels, meaning researchers are left
to devise a way to characterize projections spanning multiple voxels. Although
dMRI analysis techniques differ in many technical aspects, they generally
share two fundamental elements: quantitative estimates of fiber orientations within
each voxel and an algorithm that utilizes these estimates to find sequences of
voxels that are traversed by a putative white matter projection.}

\hl{The first step of this paradigm most commonly involves fitting a tensor
model to diffusion-weighted images \citep{Basser2000}, yielding a set of three
directional eigenvectors. The principal eigenvector serves as the estimate of
local fiber orientation. Other orientation models such as the ball-and-stick
estimated by FSL's {\tt BEDPOSTx} \citep{behrens2003characterization}, and the
coefficients of spherical harmonic basis sets estimated via constrained
spherical deconvolution (CSD) \citep{Tournier2012} are more flexible than
tensors.  Deconvolution of the diffusion orientation distribution function
(dODF) results in a ``fiber ODF'' (fODF or FOD) that is believed to be a more accurate
representation of the true fiber orientations in each voxel because the dODF
reflects fiber orientations convolved with an imaging-related point-spread
function \citep{Tournier2007,descoteaux2009deterministic}.  Other approaches
have foregone model estimation entirely and instead require a specific,
sometimes quite lengthy, dMRI scanning protocol.  For example, Diffusion
Spectrum Imaging (DSI) can calculate a dODF analytically and has been shown to
accurately represent multiple fiber orientations in a single voxel
\citep{Wedeen2005,Wedeen2008}. More recently, generalized q-sampling imaging
(GQI) can analytically calculate dODFs across most q-space sampling schemes
\citep{Yeh2010}.  The dODFs produced by GQI and DSI can be further deconvolved
or decomposed to produce fODFs. For simplicity we here refer to any function
on the sphere with regularly-spaced angles as the domain and magnitudes as the range
as an ODF, although tensors and fODF/FODs may not satisfy some other technical
definitions of an ODF.
}

\hl{After one of the aforementioned methods has been used to quantify orientations
in each voxel, the analyst can choose from a variety of algorithms that
use this information to estimate fiber connectivity between distant voxels.
Most of these tractography algorithms begin with some geometric
assumptions about the paths of axons, namely that their shapes can be
approximated by sequences of fixed-length segments that are joined at
shallow angles. \emph{Deterministic} tractography generates these sequences
by selecting seed coordinates in relevant voxels and stepping along the
directional maxima (i.e. principal eigenvectors or maxima of ODF lobes)
until a stopping criterion is met. This criterion could be that the next voxel's
peak direction(s) forms too steep an angle with the previous step
or that the voxel has been deemed un-trackable due to low SNR or an anatomical
mask. For example, consider two arbitrary regions in the brain $a,b$ that are separated by
multiple voxels. One could seed points throughout all of white matter and
generate many deterministic paths (or streamlines) and count the number
of streamlines that intersect both $a$ and $b$. This method is simple, fast,
and tends to produce biologically plausible trajectories. The method
has some limitations, however. It is difficult to interpret streamline
count \citep{Jones2013} and any projection that fans within a voxel
will not be accurately represented by the ODF peak alone.}

\hl{On the other hand, the family of so-called \emph{probabilistic} tractography
algorithms use all available orientation information; be it a full tensor, dODF
or fODF. This is accomplished by generating
paths as sequences of geometrically-compatible steps that are sampled from a
discretized version of the containing voxel's ODF as if it were a probability
mass function \citep{koch2002investigation}. Revisiting the problem of
estimating how connected region $a$ is to region $b$, the probabilistic
approach would mean seeding many points within $a$ and growing paths by
sampling the ODFs in each voxel. The number of paths that intersect $b$ out of
the total number of simulated paths serves as the estimate of connection
probability. Many popular software implementations of this approach create a
``connection density'' \citep{behrens2007probabilistic} or ``track density''
\citep{Calamante2010} image that stores a count of the trajectories that passed
through each voxel.  This count is also referred to as the Probabilistic Index
of Connectivity (PICo)\citep{parker2003probabilistic}.}

\hl{Consider the theoretical motivation behind probabilistic tractography.
In order to connect regions $a$ and $b$ we must begin
in a voxel in $a$ and continue stepping across voxels until we reach a voxel
in $b$. We are supposing that the ODF in each voxel reflects the proportion of
myelinated axons that are parallel to each direction represented in the
discretized ODF. These axons can exist anywhere in the voxel and can follow any
geometrically-allowable trajectory within that voxel. As a physical object, the
axon must enter the voxel from one of its 26 immediate neighbors and exit into
another. If we select a voxel in $a$ and seed a fixed number of points randomly
within it. All these seeds will eventually exit into a neighboring voxel and continue
to grow outwards. Only a fraction of these paths will go into each neighbor -
and a smaller fraction into the neighbor's neighbors and so on. This
has two important effects. The first is well-known: the probability of a
trajectory reaching region $b$ will decrease the more distant $b$ is from $a$.}

\hl{The second is more subtle and directly addressed by our proposed method.
In reality, some fixed proportion of axonal fibers within the seed voxel
project into each neighboring voxel. We define the proportion of axonal
fibers connecting a voxel to one of its directly adjacent neighbors as the
\emph{ground-truth transition probability} of a fiber in the seed voxel
continuing into this neighbor (if one were to randomly select an axon).
Probabilistic tractography approximates this transition probability as
the proportion of simulated paths that exit into each neighbor.
The accuracy of this approximation depends directly on the number
of seeds that are created within the voxel during probabilistic simulation. The fewer seeds
generated, the more \emph{simulation-related error} will impact the
approximation. If seed points are only generated in voxels in region $a$
this problem is compounded at each step. Fewer paths reach distant voxels,
resulting in sparser sampling of the distant voxel ODFs and greater approximation
error.}

\hl{In theory we would need an infinite number of seeds in every voxel
in order to avoid simulation-related error in connectivity estimations.
Here we introduce a method that derives closed-form equations that directly
calculate the voxel-to-voxel transition probabilities. We show experimentally that probabilistic
simulations converge to our analytically calculated values
as the number of simulated seeds increases.  Our
method precludes the need for any simulation at all, greatly decreasing the
computational demand and bypassing the invariable exponential complexity that
comes with tossing many coins at each step throughout the brain.  This increase
in efficiency does not come at the cost of any simplifying assumptions.}

\hl{Analytically calculated voxel-to-voxel transition probabilities
based on ODFs are
useful for quantifying local white matter connectivity (between a
voxel and its 26 immediate neighbors). However, additional tools are required
to capture long-range connections. One already-established method
for finding long-range projections based on local connectivity estimates
can be called Voxel Graph tractography
\citep{zalesky2008dt,iturria2007characterizing,iturria2008studying}.
This approach treats each voxel in the brain as a node in a graph
and the strength of its connectivity to its immediate neighbors as an
edge weight. The use of analytically calculated voxel-to-voxel transition
probabilities (i.e. using our closed-form solution) as edge weights in a 
Voxel Graph, defines what we call ``Analytic Tractography.''}

\hl{The
use of analytic transition probabilities provides an improvement in
accuracy, generality and speed over previously proposed Voxel Graph
methods. The original method proposed by \citet{zalesky2008dt} required
a tensor model, which will be deficient in regions of crossing, kissing
or fanning fibers \citep{Jones2013}.
Other approaches employ fast-marching methods to propagate
paths across a Voxel Graph
\citep{Campbell2005,iturria2007characterizing,sotiropoulos2010brain}.
These methods derived edge weights by constant solid angle integration
of CSD fODFs in the direction of each neighboring voxel. This constant
solid angle integration is based on the assumption that the ODF and
axons are physically centered in the voxel, which represent an 
oversimplification.
}

\hl{We present four analyses that test the validity of our method and
calculate its highest possible accuracy at capturing ground-truth transition
probabilities when using to real dMRI data. First
we show that given a set of ground-truth fibers from a software phantom,
our method can calculate ground-truth transition probabilities based on
fiber ODFs. Second, we use FiberFox to simulate dMRI from the software
phantom and estimate transition probabilities based on dMRI-derived
ODFs. Third, we empirically demonstrate that probabilistic tractography on ODFs from real
datasets converge to the analytic values as the number of seeds increases.
Finally, we show that Voxel Graph tractography based on analytic transition
probabilities produces results comparable to those from other current
methods at a fraction of the computational cost. Our demonstrations include many
current methods for ODF estimation including HARDI, generalized q-sampling
imaging, and diffusion spectrum imaging (DSI) both with and without ODF
deconvolution.}

\subsection{ODFs to transition probabilities}

We introduce two methods to calculate the probability that a white matter
structure projects into an adjacent voxel.  A closed-form expression for this
transition probability is evaluated given a voxel's ODF and a set of geometric
constraints as inputs. Geometric constraints determine the allowable complexity
of the white matter structure's shape, preventing biologically implausible
self-loops or spirals. Both methods hinge on approximating white matter
structure shape by discrete sequences of line segments called a ``turning angle
sequence''. Line segments are defined by a fixed length and orientation, with
no two sequential segments connected at an angle greater than a user-specified
fixed ``maximum turning angle''.

Crucially, the directional magnitudes in an ODF do not specify the
\emph{location} of directed axon bundles within a voxel. Turning angle
sequences accordingly do not have a fixed position within a voxel.  For
geometric constraints that prevent self-loops this ensures that the set of all
possible turning angle sequences is finite. Transition probabilities can
therefore be computed as sums over the set of turning angle sequences that
could connect a voxel to an adjacent voxel, weighted by their probabilities.
The proposed \iodfone and \iodftwo methods diverge in how these
probabilities are calculated.

For each voxel, the \emph{\iodfone} method only considers ODF values from that
voxel, assigning probabilities to turning angle sequences based on ODF
magnitudes in the directions of the turning angle sequences. By contrast, the
\emph{\iodftwo} method considers the ODFs of voxels on both sides of the
source-destination transition, ensuring that there is evidence that the
destination voxel ``agrees'' that a structure projects into it from the source
voxel.

\subsection{Fiber phantom analysis}

\hl{To test the validity of the \iodfone and \iodftwo methods, we used a set of 
artificial axonal fibers that were shaped to mimic the
layout of a coronal slice through the human brain \citep{neher2014fiberfox}.
This phantom provides an
opportunity to calculate ground truth transition probabilities by constructing a
voxel grid around the fibers and calculating the proportion of fibers that exit
into each voxel's neighbors.  We perform two sets of validation tests.  The
first uses ground-truth fODFs, where the magnitude reflects the proportion of
fibers parallel to each of a set of orientations, similar to those calculated
in Track Orientation Density Imaging \citep{dhollander2014track}. With
ground-truth fODFs as input we can compare the transition probabilities
estimated by the \iodfone and \iodftwo methods to actual ground-truth
transition probabilities in the absence of imaging-dependent noise effects on fODF
estimation. These effects are then introduced in the second set of tests, where the
ground-truth fiber dataset is simulated in the FiberFox module of the MITK
Diffusion software \citep{neher2014fiberfox}}.

\subsection{Transition probabilities in a voxel graph}

While transition probabilities are specific to a voxel and its 26 neighbors, we
can incorporate them into a Voxel Graph to trace fascicles across the
brain. The Voxel Graph is a theoretically-sound but rarely-utilized technique
described by \citet{zalesky2008dt} where white matter voxels are nodes
connected to their immediate spatial neighbors. Edge weights are based on
estimates that a white matter structure continues from one voxel into its
neighbor (i.e. transition probabilities).  By using the ``logarithm trick,''
weighting edges in the graph as the negative log of the transition
probability, one can formulate brain connectivity as a shortest path problem
\citep{zalesky2008dt}. The probability that a voxel is connected to any
other voxel is calculated as the product of the transition
probabilities along the shortest path between the voxels. The shortest path is
found efficiently using Dijkstra's algorithm \citep{dijkstra1959note}. This
forms the basis for estimating connectivity maps with analytic tractography.

To be of any practical use, analytic tractography should perform comparably to
other state-of-the-art tractography methods. To this end, we built voxel graphs
using the negative log of single-ODF and double-ODF transition probabilities
from three datasets from the human connectome project (HCP,
\citealt{VanEssen2012}). We built maps of shortest path probabilities to the
rest of the brain to demonstrate that these highlight well-known projections.
A source region in the cerebral peduncle should highlight the cortico-spinal
tract and a source region in the corpus callosum should highlight
inter-hemispheric U-fibers and lateral projections.  These projections were
reconstructed using both probabilistic (MRTRIX iFOD2 \citep{Tournier2010},
BEDPOSTx with probtrackx
\citep{behrens2003characterization,behrens2007probabilistic}, Bayesian
tractography in CAMINO \citep{friman2006bayesian}) and deterministic
tractography in DSI Studio \citep{Yeh2013deterministic}.

\subsection{Accompanying software}

This paper is accompanied by open-source software that implements all the
analyses presented here. The python package Mathematical solution for
Inter-voxel Tract Transition Expectations requiring No Simulations
(\texttt{MITTENS}) is available at github.com/mattcieslak/MITTENS.  It can
calculate transition probabilities and build and query voxel graphs.  \hl{
Although a number of groups have worked on the voxel graph method
\citep{Campbell2005,Hageman2009,iturria2007characterizing,sotiropoulos2010brain},
to our knowledge no software is publicly available that implements this method.}

\section{Methods}

Analytic tractography uses ODF values and geometric constraints to directly
calculate transition probabilities.  We present two algorithms for defining the
transition probabilities between adjacent voxels. Given a set of voxels,
$\voxels$, and a discrete set of directions $\Theta$, the diffusion probability
inside each voxel $u \in \voxels$ in \hl{each} direction $\theta \in \Theta$ is
given by $\bar{p}_{u}(\theta)$ such that $\sum_{\theta \in
\Theta}\bar{p}_u(\theta)=1$ \citep{Wedeen2005}. The number of angles in $\Theta$
is chosen by the analyst based on the angular resolution of their diffusion
scans.  While our methodology is applicable given any choice of symmetric
$\Theta$, we chose the DSI Studio default of 642 evenly spaced directions for
our presented experiments, resulting in an angular resolution of 8.09 degrees.

\subsection{Turning angle sequences}

Our approach introduces the concept of an abstract sequence of turning angles.
A turning angle sequence $\sigma = (\theta_0, \theta_{1} \hdots \theta_{n})$
has no fixed start or end points in \hl{a voxel}. It can
be thought of as a template for a path, specifying the directions taken along
the path but no spatial positioning. This abstract formulation nicely reflects
that ODFs do not specify the within-voxel location of white matter structures
\citep{kaden2007parametric,Jones2013} unlike solid angle integration methods
\citep{iturria2007characterizing,sotiropoulos2010brain} where the integration
bounds are based on the ODF being centered in each voxel.  Critically, the
turning angle sequence abstraction allows us to analytically solve for
transition probabilities between voxels.

\hl{ Each of the uncountably infinite set of potential starting positions
within the purple volume within voxel $u$ of
Figure~\ref{fig:solution_steps}(a) are equivalent in that they will result in
a transition from voxel $u$ to voxel $v$ when  following the specific turning angle sequence
$\sigma$. If turning angle sequence $\sigma$ begins anywhere else in voxel $u$
it will not end with the last hop being from $u$ into voxel $v$. 
The set of trajectories starting in the purple
volume, following $\sigma$ and ending in voxel $v$ comprise one element in the
finite set of equivalence classes that includes all possible multi-hop
trajectories that start in voxel $u$ and transition in the last hop into a
neighboring voxel.  The only necessary constraints are a maximum turning angle
$\theta_{max}$ and sufficiently large fixed step size.}

\hl{Appropriate step size and maximum turning angle parameters ensure that
each turning angle sequence eventually transitions into a neighboring
voxel, thereby ensuring the set of equivalence classes is finite. 
These constraints also prevent biologically implausible
fiber trajectories such as recurrent trajectories or loops within a voxel.
These equivalence classes enable
us to characterize all possible fiber trajectories within and between 
voxels directly.
}

\subsection{Analytic Transition Probabilities}

We quantify the probability of transitioning from voxel $u$ into
neighboring voxel $v$ by summing over the probabilities of each turning
angle sequence $(\sigma)$ that can end in $v$.  Each turning angle
sequence $\sigma = (\theta_0, \theta_{1} \hdots \theta_{n})$ has
probability given by

\begin{equation}
\Pr_u(\sigma) = \bar{p}_u(\theta_0) \times \bar{p}_u(\theta_1 \mid \theta_0)
	\times \hdots \times \bar{p}_u(\theta_n \mid \theta_{n-1})
\label{eq:sigma_probability}
\end{equation}
i.e. each probability $\Pr_u(\sigma)$ is the product of the probabilities
of taking each step in $\sigma$. The initial step $\theta_0$ has a
probability equal to the magnitude of the ODF of voxel $u$ in direction
$\theta_0$. The probabilities of subsequent hops are determined by both their
angular similarity with the preceding turning angle and their magnitude in the
ODF of voxel $u$.  For example, higher angular similarity between consecutive turning
angles could increase the probability of $\sigma$. In our implementation
we treat all geometrically compatible angles equally, and define the probability of
taking direction $\theta_i$  after taking $\theta_{i-1}$ by

\begin{equation}
\bar{p}_u(\theta_i \mid \theta_{i-1}) = \bar{p}_u(\theta_i) \mbf{1}(\theta_i,\theta_{i-1})
	\left[  \sum_{\theta' \in \Theta} \bar{p}_u(\theta') \mbf{1}(\theta',\theta_{i-1}) \right]^{-1}
\label{eq:prob_angles_weighted}
\end{equation}
where

\begin{equation}
\mbf{1}(\theta,\theta') =
\begin{cases}
	1, & \text{if the angle between~} \theta \text{~and~} \theta' \text{~is less than~} \theta_{max} \\
        0, & \text{otherwise.}
\end{cases}
\label{eq:indicator}
\end{equation}

\begin{figure}[!ht]
  \centering
    \includegraphics[height=0.6\textheight]{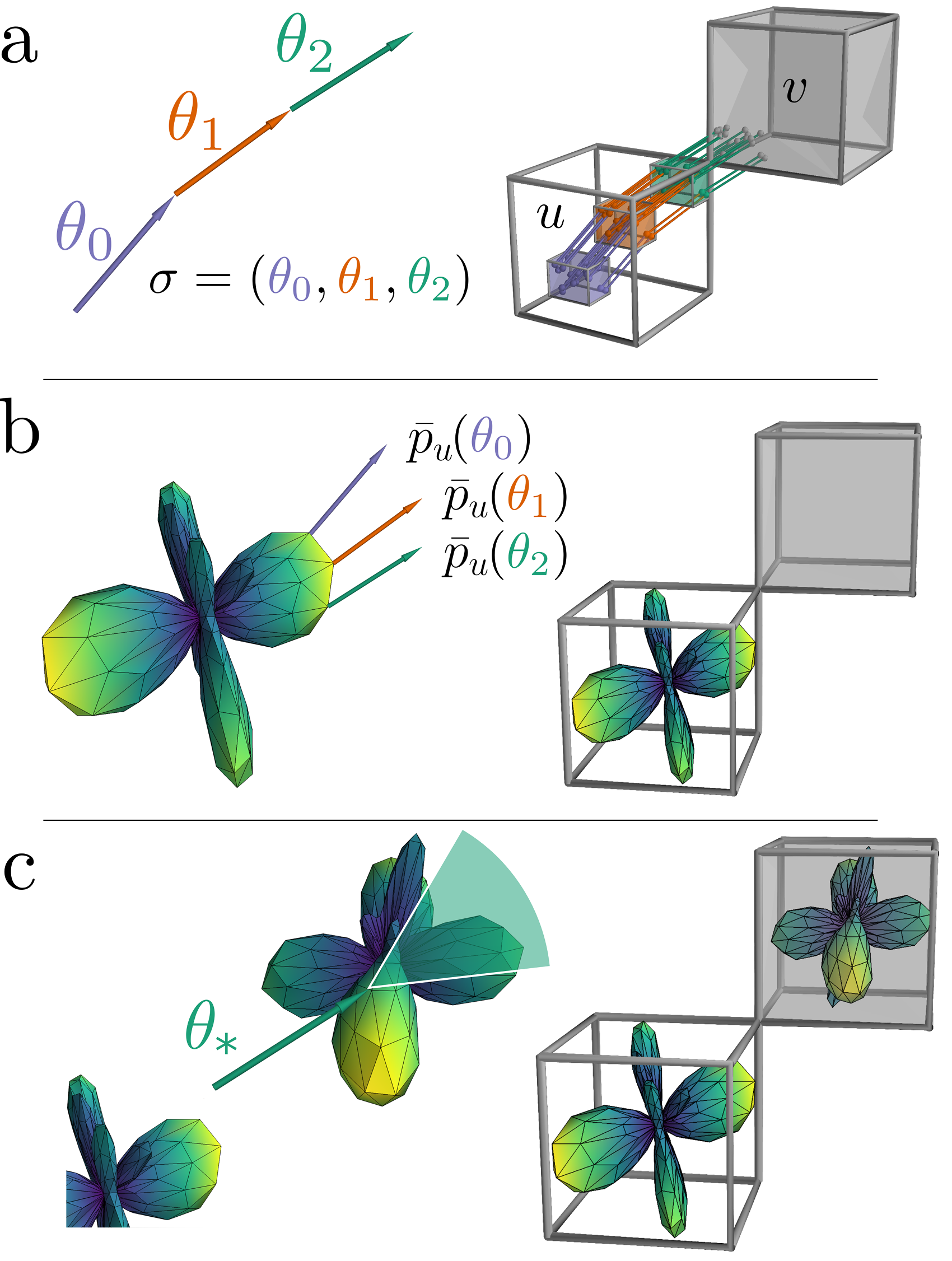}
	\caption{
		(a)
		On the left a turning angle sequence of three hops
		$\sigma=\left(\theta_0,\theta_1,\theta_2\right)$ is shown.
		This sequence will form a trajectory from source voxel $u$
		to upper-right corner voxel $v$ if it begins in the purple
		cubiod, which defines $\protect\volume\left(\sigma,v\right)$.
		A few random seeds were placed in
		$\protect\volume\left(\sigma,v\right)$ and followed $\sigma$
		through space.
		All trajectories are seen to terminate in $v$ as well as
		pass through the two intermediary volumes found by the
		recursive algorithm: $\protect\volume\left(\left(\theta_2\right),v\right)$
		is green and $\protect\volume\left(\left(\theta_1,\theta_2\right),v\right)$
		is orange.
		(b)
		A hypothetical ODF from source voxel $u$ is plotted as a
		mesh, with the distance from the center corresponding
		to the probability associated with that direction. For
		\iodfone the probabilities corresponding to
		$\theta_0,\theta_1,\theta_2$ are pulled directly from
		$u$'s ODF.
		(c)
		Transition probabilities calculated by \iodftwo
		incorporate the ODF from voxel $v$. The directions
		in $v$'s ODF compatible with the direction
		taken into $v$, $\theta_*=\theta_2$ in this illustration, are included
		in the transition probability calculation.
		}
\label{fig:solution_steps}
\end{figure}

For a given $\sigma$ and a neighboring target voxel $v$, we define
$\volume(\sigma,v)$ as the volume given by integrating over all the potential
starting points for $\sigma$ in voxel $u$ that result in the trajectory
terminating in $v$ without first stepping into any other voxel (see
Figure~\ref{fig:solution_steps} for an example).  We implement a recursive
algorithm to find the complete set of turning angle sequences $\mathit{\Sigma}$
and their corresponding volumes, $\volume$.

\subsection{Recursive algorithm}

Consider a starting voxel $u$ and a neighboring target voxel $v$. A one-hop turning
angle sequence from $u$ to $v$ is any $\sigma = (\theta_*)$ for which
there exists $x,y$ such that $x \in u, y \in v,$ and $y = x + s\theta_*$
where $s$ is the step size. We begin by finding all possible one-hop
turning angle sequences that start in $u$ and end in $v$.  For each
one-hop trajectory there is a volume $\volume((\theta_*),v)$ in $u$ where
the trajectory can begin such that it ends in $v$. Of note, this
exhausts all possible one-hop trajectories from $u$ to $v$ for the given
step size. All possible two-hop trajectories must be extensions of
one-hop trajectories that start in $u$. In other words, two-hop turning
angle sequences are defined \emph{backwards}, by finding any possible
hop within starting voxel $u$ that ends in $\volume((\theta_*),v)$ for some
one-hop turning angle sequence $(\theta_*)$. Further backwards hops are
added until no more compatible hops within $u$ can be found. In summary,
this algorithm exhausts all possible turning angle sequences from $u$ to
$v$ by first obtaining the critical $\volume((\theta_*),v)$ from which a
single hop can terminate in target voxel $v$ and then working backwards
to identify all turning angle sequences and associated volumes. The
resulting set of multi-hop turning angle sequences originating in
starting voxel $u$ is complete because all turning angle sequences of
length $n$ must be extensions of turning angle sequences of length $n-1$.

For example, the trajectory $\sigma$ illustrated in
Figure~\ref{fig:solution_steps} was obtained by first determining the
volume (green cuboid) in $u$ from which a hop in direction $\theta_* = \theta_2$
terminated in target voxel $v$.  Moving backwards, the next volume
(orange cuboid) was then defined such that a hop in direction $\theta_1$
terminated in the green cuboid. Repeating this step for $\theta_0$
produced the purple cuboid, after which no further hops were possible
within $u$ for any $\theta$ compatible with $\theta_0$.

Without biologically reasonable constraints on step size and maximum
turning angle, one cannot guarantee that the above algorithm terminates.
A large angular threshold or a small step size could result in infinite
recursion. However such parameters will also lead to biologically
implausible structures such as recurrent loops and spirals within a
voxel. We use a fixed step size of $\sqrt{3}/2$ voxels and a turning
angle maximum of 35 degrees.

Importantly, the set of turning angle sequences is independent of empirical
input data and only needs to be run once per step size and maximum turning
angle.  In practical terms, a set of turning angle sequences is calculated once
(already included in the accompanying \texttt{MITTENS} software) and can
subsequently be applied to any number of empirical datasets without the need
for recalculation.  In the following we define two complimentary methods that
utilize the precalculated set of turning angle sequences to estimate transition
probabilities on empirical data.

\subsection{The single-ODF approach}

The goal of the \iodfone approach is to define the transition probability
between each voxel $u$ and each of its adjacent voxels for empirical data.  For a
given voxel $v$ from the 26 neighbors of $u$ $(\mathcal{N}(u))$, this
transition probability from $u$ to $v$ is given by:

\begin{equation}
	\Pr(u \rightarrow v) = \sum_{\sigma \in \mathit{\Sigma}}\Pr_u(\sigma) \volume( \sigma,v )
\label{eq:transition_prob}
\end{equation}
where $\mathit{\Sigma}$ is the finite set of turning angle sequences and
$\volume( \sigma,v )$ is a proportion of the total voxel volume 
($\volume( \sigma,v ) \in [0,1]$).

In this approach the transition probabilities depend only on the ODF of the
starting voxel $u$. As illustrated in Figure~\ref{fig:solution_steps}b,
$\Pr_u(\sigma)$ is completely determined by the ODF in $u$. The
\iodfone transition probabilities from $u$ to each of its  26 adjacent voxels
are given by equation~\ref{eq:transition_prob}, concretized by the ODF of the
starting voxel $u$. At a high level, the assumption that white matter structures
influence the ODF motivates our use of the ODF to estimate the probabilities
of existing white matter structures projecting through each starting voxel into its
neighbors. The \iodfone method elegantly transforms the starting voxel's ODF
information into estimated transition probabilities.

\subsection{ The double-ODF approach }

While the \iodfone approach produces a complete map of transition probabilities
for the entire brain, it also introduces a potential problem. It computes a
non-empty set of turning angle sequences connecting a voxel to all its
neighbors that, on first glimpse implies that all voxels contain white matter
structures connecting them to their neighbors. This is biologically
implausible.  To reduce this problem we introduce the
\iodftwo method, which simultaneously considers ODFs in both the starting and
the target neighboring voxel. By penalizing conflicting directional diffusion information
between ODFs of neighboring voxels, the transition probabilities are lower for
neighbors that do not have compatible ODF values that provide evidence of white
matter connections.

Our goal is to adjust transition probabilities in favor of trajectories likely
to continue after exiting the starting voxel. To do so, the \iodftwo method
adjusts the probabilities of the turning angle sequences based on the ODF in
target voxel $v$. Intuitively, the probability of a turning angle sequence
exiting $u$ at angle $\theta_*$ will be increased if directions compatible with
$\theta_*$ have relatively high probability within $v$'s ODF.  This is reflected
mathematically through the introduction of a weighting term
(\ref{eq:weighting}) that uses $v$'s ODF to calculate a similarity-based weight for
a next step compatible with $\sigma$ existing in $v$:

\begin{equation}
	w(\sigma,v) = \sum_{\theta} \bar{p}_v(\theta) \mbf{1}(\theta,\theta')
	\label{eq:weighting}
\end{equation}
where $\bar{p}_v(\theta)$ is the ODF value for $\theta$ in
$v$, and $\mbf{1}(\theta,\theta_*)$ is defined in equation~\ref{eq:indicator}.
The summation is finite since $\Theta$ is finite.

We use conditioning on $(u\rightarrow \bullet)$ to denote the assumption that a
``compatible'' next hop exists in \emph{at least one} neighbor of $u$ after a
turning angle sequence exits $u$.  The \iodftwo method computes the transition
probabilities:

\begin{equation}
\Pr\left(u \rightarrow v \mid u \rightarrow \bullet \right) = \frac{1}{\alpha}
\sum_{\sigma \in \mathit{\Sigma}} \Pr_u\left(\sigma\right)\volume\left(\sigma,v\right)w(\sigma,v)
\label{eq:iodftwo}
\end{equation}
where
\begin{equation}
\alpha =
	\Pr(u \rightarrow \bullet) = \sum_{v\in\mathcal{N}(u)} \Pr(u \rightarrow \bullet | u \rightarrow v) \Pr(u \rightarrow v)
\end{equation}

We derive equation \ref{eq:iodftwo} using Bayes Theorem, under the
assumption that the starting point for turning angle sequence $\sigma$
(moving from $u$ to $v$) is uniformly distributed within $\volume(\sigma,v)$:

\begin{eqnarray*}
	\lefteqn{  \Pr(u \rightarrow v \mid u \rightarrow \bullet) } \\
 &=&  \frac{\Pr(u \rightarrow \bullet  \mid  u \rightarrow v) \Pr(u \rightarrow v)}{\Pr(u \rightarrow \bullet)}\\
	&=& \frac{1}{\alpha} \sum_{\sigma \in \mathit{\Sigma}} \Pr(u \rightarrow \bullet \mid \sigma \text{~and~} u \rightarrow v)  \Pr(\sigma \mid  u \rightarrow v)\volume\left(\sigma,v\right) \Pr(u \rightarrow v)\\
	&=& \frac{1}{\alpha}\sum_{\sigma \in \mathit{\Sigma}} w(\sigma,v)\volume\left(\sigma,v\right) \Pr(\sigma \mid u \rightarrow v)Pr(u \rightarrow v)\\
	&=& \frac{1}{\alpha}\sum_{\sigma \in \mathit{\Sigma}} w(\sigma,v)\volume\left(\sigma,v\right) \frac{\Pr_u(\sigma)}{\Pr(u \rightarrow v)}
                         \Pr(u \rightarrow v)\\
	&=& \frac{1}{\alpha}\sum_{\sigma \in \mathit{\Sigma}} w(\sigma,v) \volume\left(\sigma,v\right)\Pr_u(\sigma) \\
\end{eqnarray*}

In contrast to the \iodfone version, the \iodftwo approach is not purely local
but depends on both the ODF in the current voxel and that in the target
neighbor. As a result, the relative probability of a trajectory out of $u$ into
a neighbor increases for those neighbors that are more likely to contain a
compatible trajectory with which to continue the path.

The \iodftwo approach is comparable to the iFOD2 algorithm \citep{Tournier2010}
included in the MRTRIX software package in that they both utilize ODFs (or
FODs) in both the source and destination voxels while tracking.  However, they
differ in that iFOD2 re-combines \hl{(by interpolation)} the source and destination FODs \emph{during}
tracking. As it is currently formulated, the iFOD2 approach does not have an
analytic solution so must be simulated.

\hl{The \iodftwo method is noteworthy in that it can result in a set of
transition probabilities that are asymmetric about the center voxel.
This could be a particularly useful feature in voxels where a y-split or
fanning is not well-represented by a symmetric ODF. Recently
methods for generating asymmetric ODFs based on surrounding voxels 
have been introduced \citep{reisert2012geometry,bastiani2017improved}.
The ability to capture asymmetric
fiber populations within voxels was shown to reliably capture 
fiber patterns commonly seen histologically \citep{bastiani2017improved}.
These methods  fit hyperparameters that introduce asymmetry to
fODFs based on neighboring voxels.
The \iodftwo method does not require any specific ODF reconstruction
technique and does not fit any parameters. There is also no theoretical barrier to using asymmetric ODFs as input to
\iodfone or \iodftwo equations.}

\subsection{Ground-truth validation test }
\hl{
The critical validity test for the \iodfone or \iodftwo methods is whether
they accurately calculate ground-truth transition probabilities when the
underlying fibers are already known. This is made possible by the FiberFox
\citep{neher2014fiberfox} software included in the MITK Diffusion package.  FiberFox
takes as input a set of 3D ``fibers'' that represent myelinated axons. These
fibers generate axon-like signal in simulated dMRI volumes, which can then
be used to test whether a specific dMRI analysis pipeline is accurate in
recreating the input fibers.}

\begin{figure}[!ht]
  \centering
    \includegraphics[width=1.0\textwidth]{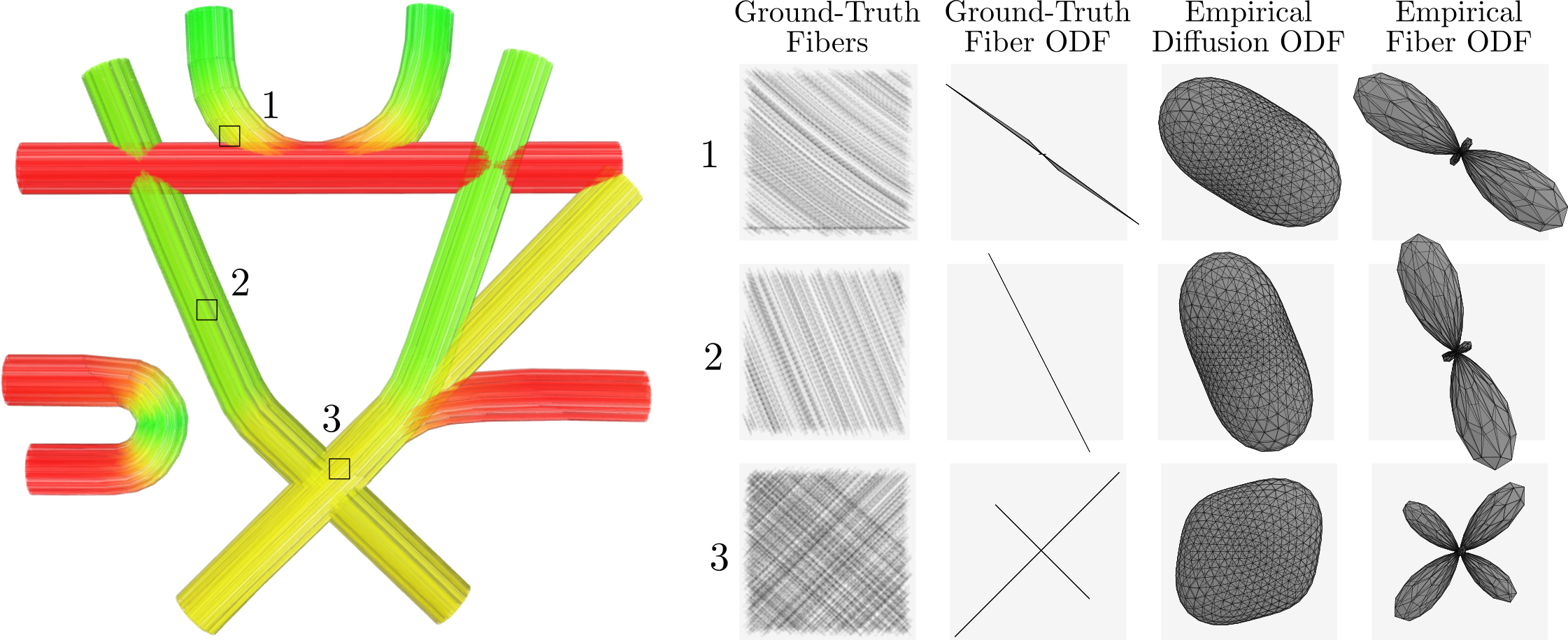}
	\caption{
        \hl{
        Left panel: The ground-truth fibers depicted as 3D streamlines.
        Three ``voxels'' were chosen to illustrate the differences between
        ODF types.
        Right panel: Each of the three illustrative voxels are displayed
        as rows of plots. The Ground-Truth Fibers column plots all
        ground-truth fiber segments contained within the voxel. The Ground-Truth
        Fiber ODF column shows the distribution of 0.1mm fiber segments
        along the canonical set of 642 directions. These same voxels
        were simulated in FiberFox using the HCP protocol. The GQI
        dODF and deconvolved fODF are shown in the final two columns.}
        }
\label{fig:ground_truth}
\end{figure}

\hl{We used the public release of the FiberCup fiber phantom from
nitrc.org.  The included fiber trajectories follow the shape of a
well-known physical phantom
\citep{poupon2010diffusion,fillard2011quantitative} designed to mimic a
coronal slice through the human brain (Figure~\ref{fig:ground_truth}, left).
Synthetic voxel grids were created covering the spatial extent of the fibers.
At each voxel we calculated the ground-truth transition probabilities as the
proportion of ground-truth fibers that intersect each of the 26 neighboring
voxels. Within each voxel we also calculated a ground-truth fODF by breaking
the contained fiber trajectories into 0.1mm segments. An angular histogram
of these segments was calculated such that the histogram bins matched
the angles represented in a 642-direction ODF. To match the symmetry in ODFs, 
counts in each bin were
made symmetric by adding the counts along mirrored directions and normalized
into probabilities.  These ground-truth fODFs were used as input to the
\iodfone and \iodftwo equations and our closed form outputs were compared to the
corresponding ground-truth transition probabilities.}

\hl{No method currently exists to compute exact fODFs in living human brains,
leaving us with imperfect estimations based on dMRI data.  Indeed, we see in
Figure~\ref{fig:ground_truth} that the second column in the right panel depicts
ground-truth fODFs that are mostly delta functions.  This indicates no
uncertainty or spread in the direction of fibers in a voxel, which would never
be observed in dMRI-based ODFs.  We therefore also used FiberFox
\citep{neher2014fiberfox} to simulate realistic dMRI volumes based on the
ground-truth fibers sampled with the 270-direction multi-shell HARDI sequence
from the human connectome project to create more realistic approximations of
dMRI data, based on the ground-truth phantom structure. These simulated dMRI
volumes were reconstructed in DSI Studio using GQI both with
(Figure~\ref{fig:ground_truth}, fourth column) and without deconvolution
(Figure~\ref{fig:ground_truth}, third column). The difference between analytic
transition probabilities based on dMRI-derived ODFs provides a more realistic
estimate of our method's performance with empirical data.}

\subsection{Comparison of analytic and probabilistic transition probabilities}

\hl{Next we selected ODFs from two human dMRI datasets to empirically assess whether
probabilistic simulations converge to the analytic transition probabilities
as the number of seeds increases.}
We applied the \iodfone and \iodftwo methods to two high-quality dMRI datasets.
Both were reconstructed with multiple ODF reconstruction methods to evaluate
the stability of calculated transition probabilities across a range of popular
preprocessing pipelines.  

The first data set was a multishell HARDI scan from a randomly
selected subject from the human connectome project 900 subject release.  The
data were downloaded from the minimal-preprocessing repository
(connectomedb.org).  Diffusion scans were acquired at a spatial resolution
$1.25\times1.25\times1.25$mm, with three q-shells at b = 1000, 2000, and 3000
s/mm$^2$, 90 directions and 10 b0 images per shell. The minimal preprocessing
pipeline included eddy current and motion correction, gradient unwarping, and
geometric distortion correction using information from acquisitions in opposite
phase-encoding directions \citep{glasser2013minimal}. These images were
reconstructed in DSI Studio \citep{Yeh2010} using GQI, with ODF
deconvolution (regularization parameter set to 0.5), CSD (harmonic order=6,
automated point spread function estimation, QBI regularization set to 0.006)
and CSD with deconvolution (same parameters).

The second dataset included a total of 709 directions on a 7th-order Cartesian
grid with a maximum b-value of 5000 s/mm$^2$ and voxel size of
$1.8\times1.8\times1.8$mm. A total of 20 additional b0 images were acquired at
regular intervals during the scanning.  These images were extracted and used to
perform motion correction. This dataset was reconstructed with both the original
DSI technique (smoothing window length of 16) and GQI, both with and without
ODF deconvolution (regularization parameter set to 0.5). \hl{No preprocessing was 
performed on the DSI data.}

\begin{figure}[!ht]
  \centering
    \includegraphics[width=0.8\textwidth]{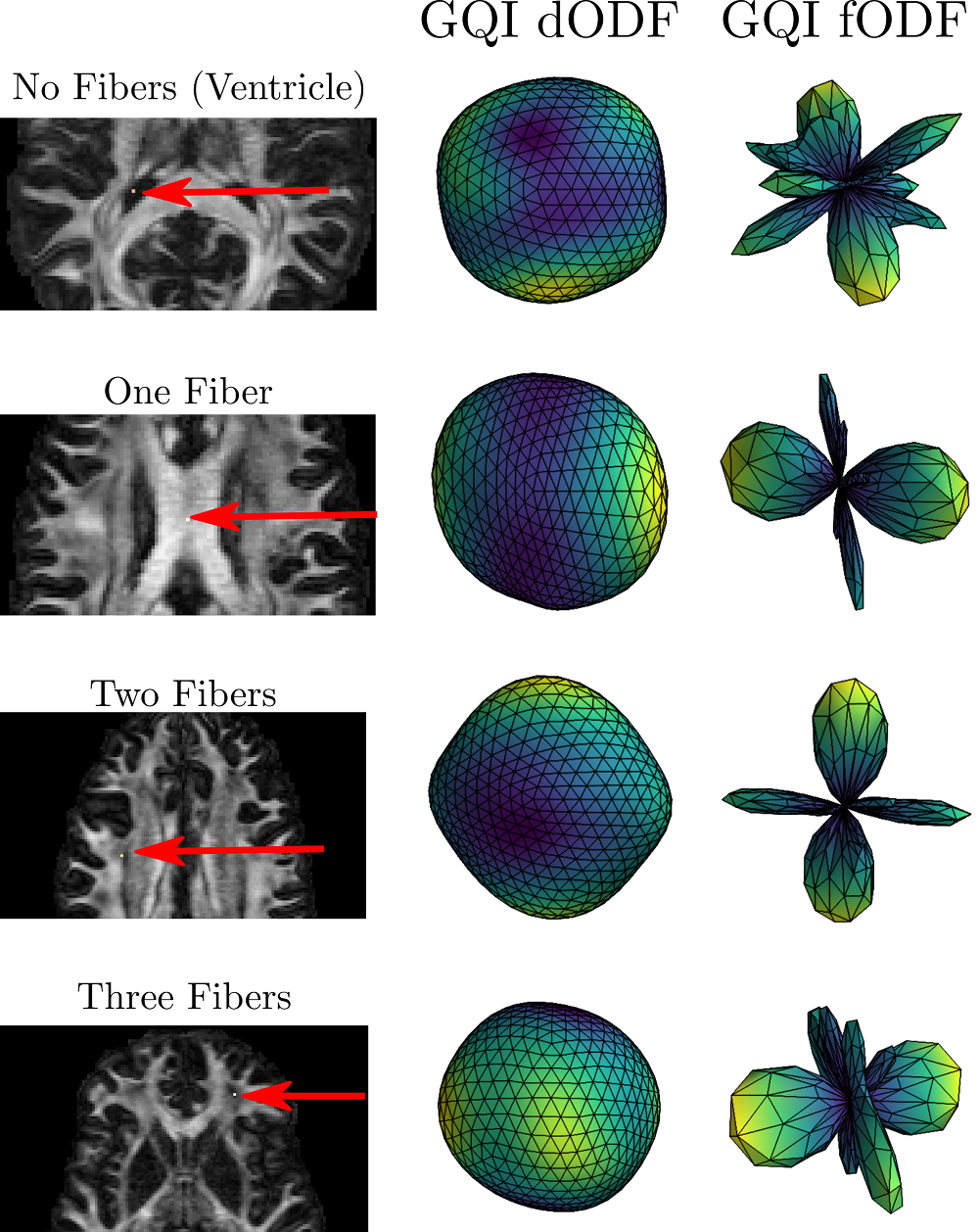}
  \caption{\hl{A single voxel was selected from four different regions
  of the brain, each with a different expected number of fiber populations.
  The source voxels are plotted in their containing axial slices
        (left column). The GQI diffusion ODF (dODF) from that voxel is shown
        in the middle column and the deconvolved GQI fiber ODF (fODF) in the
  right column.}}
\label{fig:source_voxels}
\end{figure}

Four voxels were selected from each dataset to use for both probabilistic
simulations and analytic transition probability calculation (Figure~\ref{fig:source_voxels}).
One was located in
the middle of the third ventricle (such that it contained no choroid plexus).
This voxel contains freely diffusing cerebrospinal fluid and no white matter
structures. Another voxel was chosen from the middle of the corpus callosum,
which contains a fiber population projecting in a single direction. The third
voxel came from a voxel in frontal lobe white matter that contains fibers from
a local U-fiber and the superior longitudinal fasciculus. The fourth voxel
contained fibers from the corticospinal tract, superior longitudinal fasciculus
and a local U-fiber.  ODFs in each of these four voxel types from both DSI and
HCP datasets were estimated using all four reconstruction methods. To contrast
the accuracy of our analytic, simulation-free method with the variability
of transition probabilities estimated via simulations, we implemented the
stochastic sampling algorithm of \citet{koch2002investigation} and calculated
the distribution of transition probability estimates from 1000 runs of
simulating $10^3$, $10^4$, $10^5$, and $10^6$ seeds in each voxel. The means of
these simulation-estimated transition probability distributions, \hl{based on
real ODFs}, were compared
to the corresponding analytic transition probabilities calculated using the
\iodfone method.

\subsection{Comparison of analytic voxel graphs to current tractography methods}

Our goal here is to show that analytic transition probabilities can be used in
a voxel graph to track known fascicles.   We randomly selected three additional
subjects from the HCP 900 subject release on which to run probabilistic,
deterministic and analytic tractography.  A multimodal symmetric group template
\citep{Avants2007,avants2010optimal} was created with a 1.25mm isotropic voxel
size from T1-weighted, T2-weighted and GFA volumes from a 28-subject age,
gender and ethnicity stratified subset of the HCP 900 subject release.  Our
three selected subjects' T1-weighted, T2-weighted and GFA volumes were
normalized to these templates using the SyN algorithm with cross correlation
\citep{avants2008symmetric}.

\begin{figure}
\begin{minipage}{\textwidth}
\subsection*{Source regions used for tractography}
\centering
\includegraphics[width=1.1\textwidth]{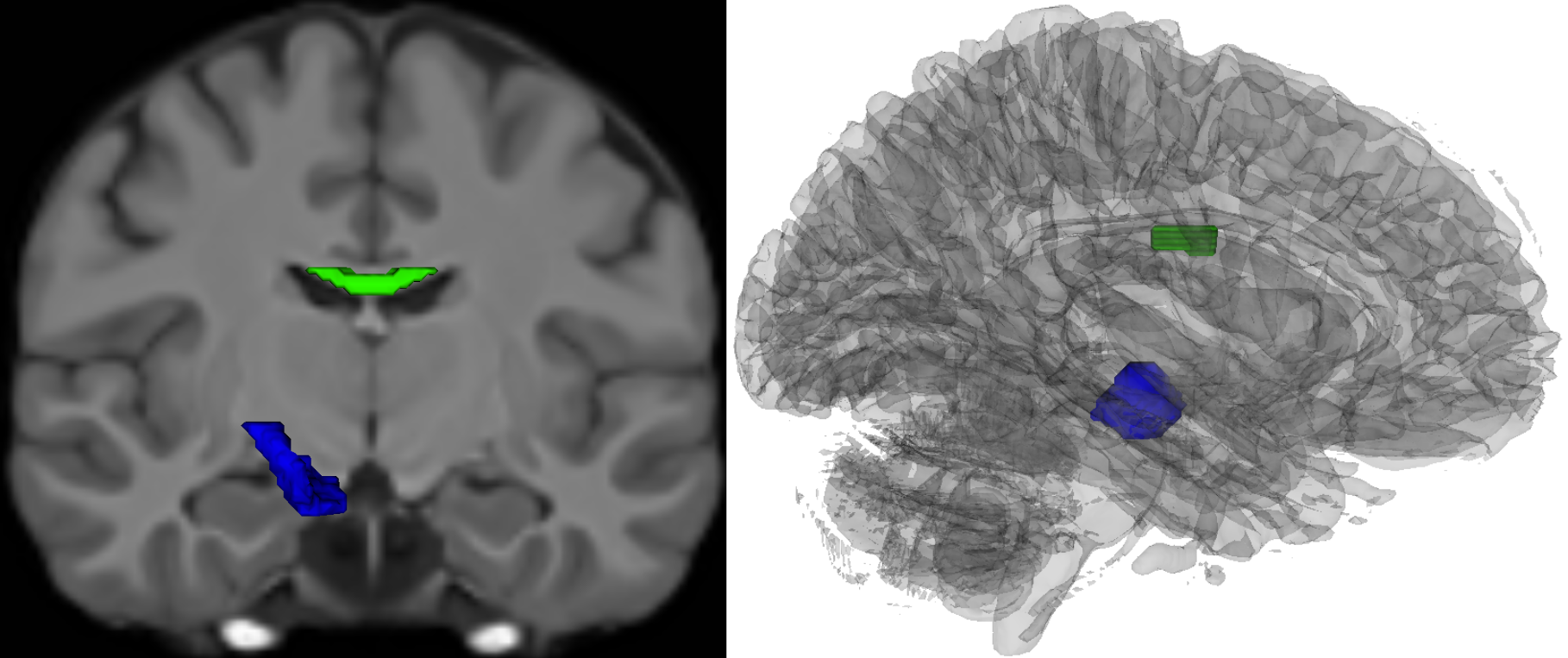}
\caption{The corpus callosum source region (green) and cerebral peduncle
		regions are plotted along with the T1-weighted image from the group template.
		Left shows the regions as 3D surfaces and right shows the regions
		inside a 3D rendering of the template brain.}
\label{fig:source_regions}
\end{minipage}
\end{figure}

Source regions were drawn on the template image in the right cerebral peduncle
and corpus callosum (Figure~\ref{fig:source_regions}).  The source regions were
warped into each subject's native space for tractography. The \hl{voxelwise connectivity maps}
from each tractography method were warped into template space for comparison. The
runtime of each tractography method was also recorded.

\paragraph{Analytic tractography}

These were processed using GQI reconstruction \hl{with fiber
deconvolution} in DSI Studio. Single-ODF and double-ODF transition probabilities
were calculated in {\tt MITTENS} and voxel graphs were built using both
formulations. Step size was $s=\sqrt{3}/2$ and turning angle max was $\theta_{max}=35$ degrees.
Adopting the approach taken by \citet{zalesky2008dt}, we define
the probability of a path as the product of the probability
of each step along the path. By transforming each transition
probability by the negative logarithm we ensure that the additive
objective used in Dijkstra's algorithm is equivalent to maximizing the
product of along-path probabilities\footnote{{\tt MITTENS} supports other metrics
for finding paths between voxels such as the ``bottleneck'' metric used by
\citealt{iturria2007characterizing}.}. Since these along-path probabilities
are greatly attenuated with each additional step in the path, we quantify
the strength of connection between two voxels as the mean probability along
the connecting path. Finally, to generate the probability map, we assign
to each voxel the maximum probability of any shortest path going through it.

It is worthwhile to note that
this approach can potentially be improved by including curvature constraints
during the path finding \cite{sotiropoulos2010brain}.
We use Dijkstra's algorithm directly without a curvature constraint to establish
a lower bound of performance.

\paragraph{Deterministic: DSI Studio}
The same GQI reconstructions used for analytic tractography were used for
deterministic tractography in DSI Studio. We again used a step size of
$\sqrt{3}/2$, a turning angle maximum of 35 degrees, a minimum length of 10mm,
maximum length of 140mm, and a QA threshold automatically determined by
DSI Studio. All white matter voxels were seeded and tracking was performed
until $10^6$ streamlines intersecting the source region were reconstructed.
This set of streamlines was used to create a track density image that was used
as the summary output for this method.

\paragraph{Probabilistic: MRTRIX}
These same preprocessed datasets were processed in MRTRIX3. Multi-shell
multi-tissue constrained spherical deconvolution was used to reconstruct fODFs
in each voxel, lmax = 8 \citep{jeurissen2014multi}. Probabilistic tractography
was run using the iFOD2 algorithm with anatomically constrained tractography
enabled and the following parameters: step size of 0.5mm, maximum length of
250mm, maximum angle between steps 80 degrees, 10,000 randomly placed seeds per voxel.
The output tracking was cropped at the GM-WM interface and converted into a
track density image for comparison to other methods.

The tracking output
was summarized as a track density image, which we used for comparison to other
methods.

\paragraph{Probabilistic: CAMINO}
The datasets were processed in CAMINO so that Bayesian probabilistic
tractography could be performed \citep{friman2006bayesian}. We followed the
recommended steps on the CAMINO website for processing HCP data.  Diffusion
weighted images where $b>1000$ were discarded before passing data to CAMINO's
{\tt track} program. The Bayes Dirac  model was used with a curvature threshold
of 70 degrees over 5mm and the reduced set of 1082 directions for 1000
iterations.  Tracking results were here also summarized as track density images.

\paragraph{Probabilistic: probtrackx}
The results of processing these same subjects with {\tt BEDPOSTx} were
downloaded from the HCP database.  These were further processed with {\tt
probtrackx2} using the parameters recommended for HCP analysis in the FSL
mailing list. Specifically, loop checking was enabled, distance correction was
applied, default curvature threshold of 0.2, default volume fraction of 0.01,
2000 step maximum, 10,000 samples, step size of 0.5mm, voxel center seeding,
and no distance threshold. The corrected output path distribution was used for
comparison to other methods.

\section{Results}

\subsection{\hl{Analytic transition probabilities vs ground-truth transition probabilities}}

\begin{figure}[!ht]
  \centering
    \includegraphics[width=1.0\textwidth]{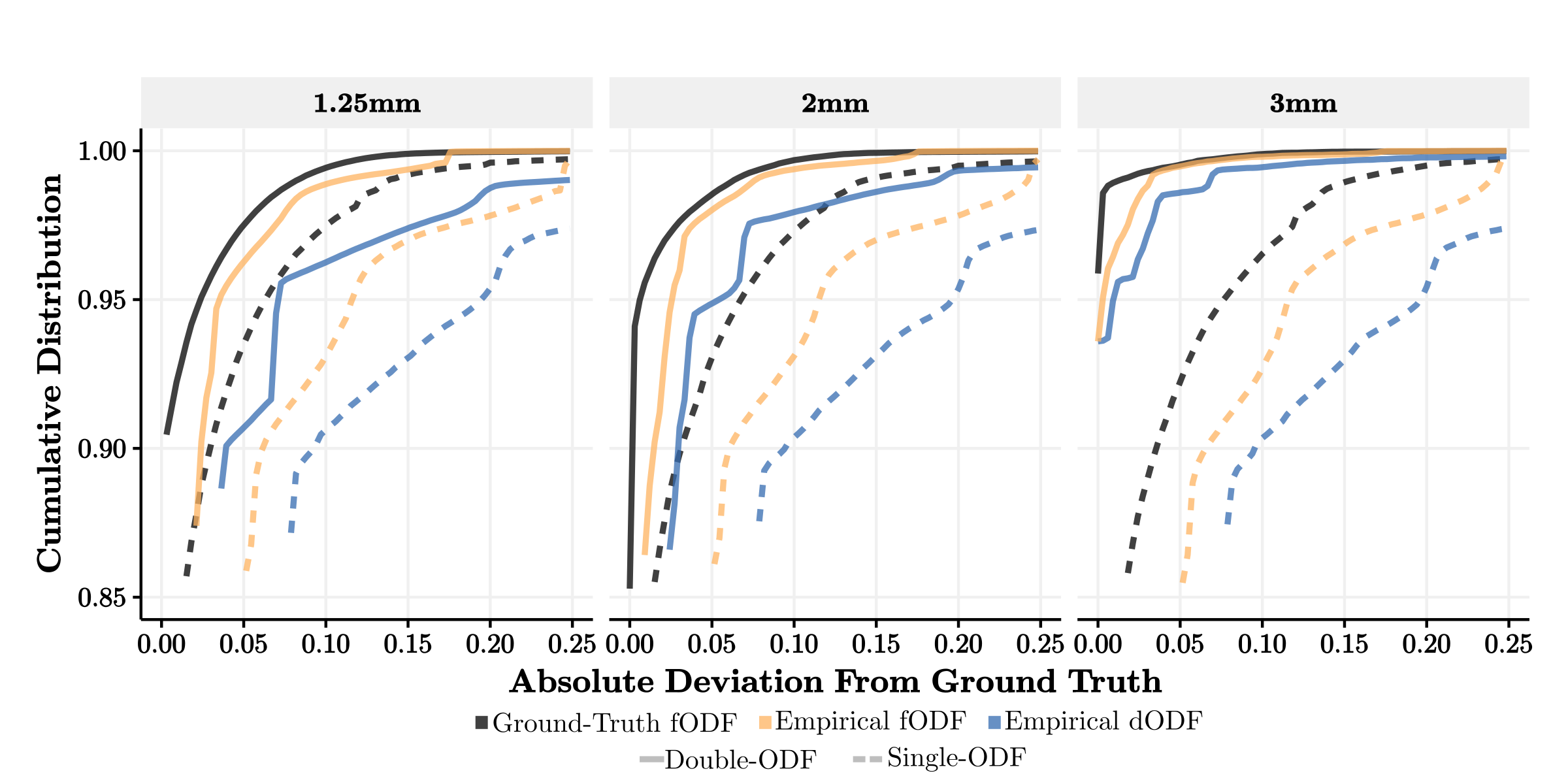}
	\caption{
        \hl{This figure plots the error CDF of analytic transition
        probabilities with three different types of ODFs as input.  The results
        for ground-truth fODFs are plotted in black, the others were ODFs
        reconstructed from FiberFox simulated dMRI: GQI dODFs are plotted in
        blue and deconvolved GQI fODFs in yellow.  The error CDFs for \iodftwo
        transition probabilities are solid lines and \iodfone are dashed. Only
        voxels with 26 neighbors inside the fiber phantom were included. 
        The lines begin at the percentile of the first non-zero error observed.
        Results for 1.25mm, 2mm and 3mm voxels are presented in separate 
        panels.}
        }
\label{fig:ground_truth_errors}
\end{figure}

\hl{ Figure~\ref{fig:ground_truth_errors} shows the empirical cumulative
distribution function (CDF) of absolute differences between analytic transition
probabilities and ground-truth transition probabilities for \iodfone and
\iodftwo calculations based on each of three types of input ODF values (6
curves per panel), for each of three voxel resolutions (three panels). 
The \iodftwo method was more accurate than \iodfone overall.}

\hl{At first glance these CDFs seem to indicate that there is considerable
error -- particularly for \iodfone transition probabilities. It is important to
note that these are CDFs for the absolute value of the error. For \iodftwo the
least accurate condition was 1.25mm voxels with a mean error of $10^{-20}\pm
0.04$ standard deviation. For \iodfone the least accurate condition was 3mm
dODF with a mean error of $5.4^{-11}\pm 0.08$ standard deviation. Recall
that errors can be introduced by the forced-symmetry or the discretized 
angles in the ground-truth fODFs. The same step size and maximum turning 
angle were used for each voxel size, which means that different 
across-voxel fiber trajectories were possible at each resolution. This
likely also contributes to error rate.}

\hl{Critically, the face-validity check (using ground-truth fODFs) of the
analytic transition probabilities against ground-truth transition probabilities
shows that the errors at the 95th percentiles are quite low. For the \iodftwo
method the errors at the 95th percentile were all 0 at 3mm, $\leq0.006$ at 2mm,
and $\leq0.02$ at 1.25mm. For the \iodfone method the errors at the 95th
percentile were $\leq0.08$ at 3mm, $\leq0.07$ at 2mm, and $\leq0.06$ at 1.25mm.
The \iodftwo method is notably more successful at recovering the true
transition probabilities.}

\subsection{Probabilistic transition probabilities converge to analytic values}

\begin{figure}[!ht]
  \centering
    \includegraphics[width=1.0\textwidth]{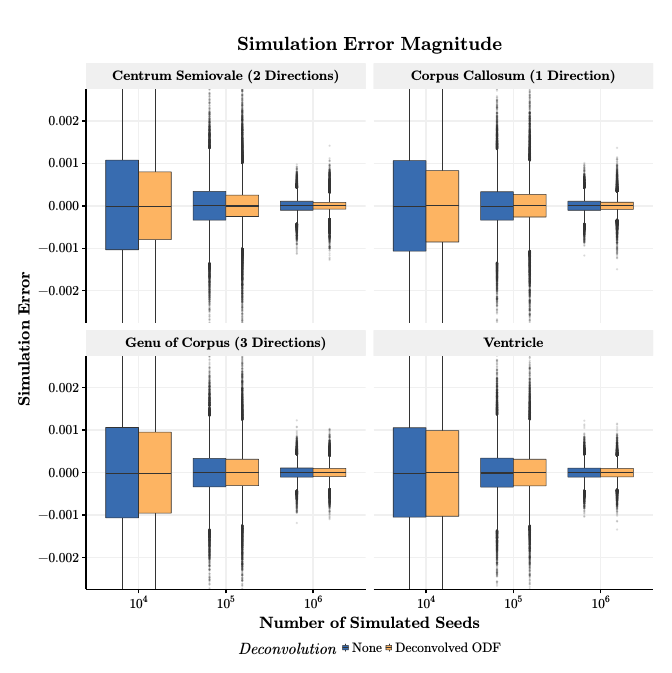}
  \caption{\hl{Distributions of the magnitude of simulation-related error as a
        function of the number of simulated seeds. Depicted in the boxplots are
        all differences between the simulation-based transition probability
        estimation and the analytic transition probability using the \iodfone
        method. Each panel depicts a voxel from Figure~\ref{fig:source_voxels}.
        ODFs used here came from an HCP HARDI dataset reconstructed using GQI.
        Blue boxes display the dODF-based and yellow display the fODF-based error
        distributions.}}
\label{fig:gqi_error_distributions}
\end{figure}

Figure~\ref{fig:gqi_error_distributions} empirically demonstrates that
probabilistic simulation-generated transition probability estimates converge to
the analytic transition probabilities when operating on the same ODF data. The
differences between the simulation distribution means and analytic transition
probabilities were less than $10^{-10}$ when $10^{4}$ or more seeds were
simulated. This is notable since $10^{-10}$ is smaller than IEEE float32
precision (which is accurate up to $10^{-6}$) but still representable by the
float64 datatype used during the simulation (which can represent up to
$10^{-15}$).

\hl{Note that the variability of the stochastic simulation estimates increases
when fewer seeds are sampled. For example, simulation-related error
distributions plotted in Figure~\ref{fig:gqi_error_distributions} reflect that
the estimates from $10^6$ seeds have substantially lower error ($\pm10^{-5}$)
compared to those using $10^4$ seeds ($\pm10^{-4}$).}

\begin{figure}[!ht]
  \centering
    \includegraphics[width=1.0\textwidth]{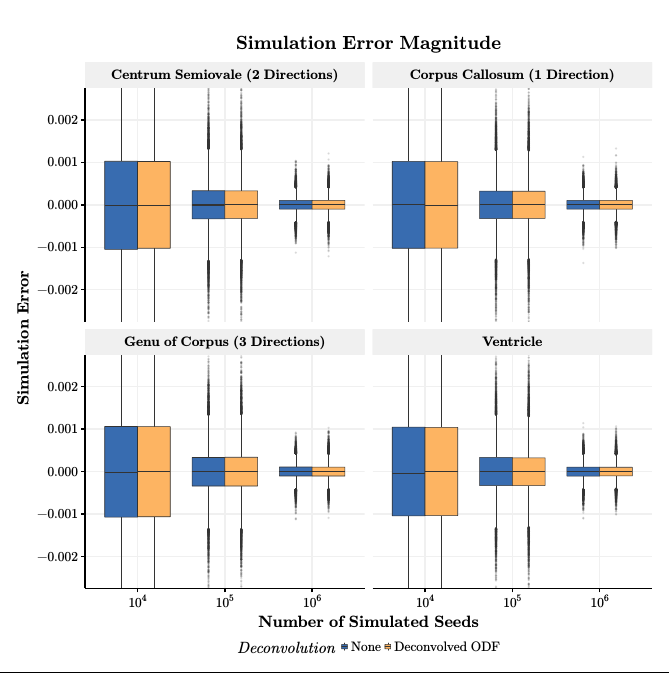}
  \caption{\hl{Distributions of the magnitude of simulation-related
  error as a function of the number of simulated seeds. Each panel
  depicts a voxel from Figure~\ref{fig:source_voxels}. ODFs came from
  a high-quality DSI dataset.}}
\label{fig:dsi_error_distributions}
\end{figure}

Analytic transition probabilities are accurate regardless of how the data were
acquired (DSI or multishell HARDI), how the ODFs were reconstructed (GQI, DSI
and CSD), whether or not ODF deconvolution was applied and regardless of the
complexity of the tissue contained in the voxel. The results for the 7th-order
Cartesian grid q-space sampling (i.e. a high-quality DSI scan) are presented in
figure~\ref{fig:dsi_error_distributions}.

Notably, in both the HCP and DSI cases the simulation means begin to converge
to the analytic solution when $10^4$ or more seeds per voxel are simulated.
However even $10^4$ seeds do not provide transition probability estimates that
are free from simulation error. The simulation-related error in transition
probabilities decreases by a factor of 10 for each corresponding increase by a
factor of 10~in the number of simulated seeds, whereas our analytic results are
free of simulation-related error.

\subsection{Analytic voxel graph tractography is fast and comparable to other methods}

\begin{figure}[!ht]
  \centering
    \includegraphics[height=0.7\textheight]{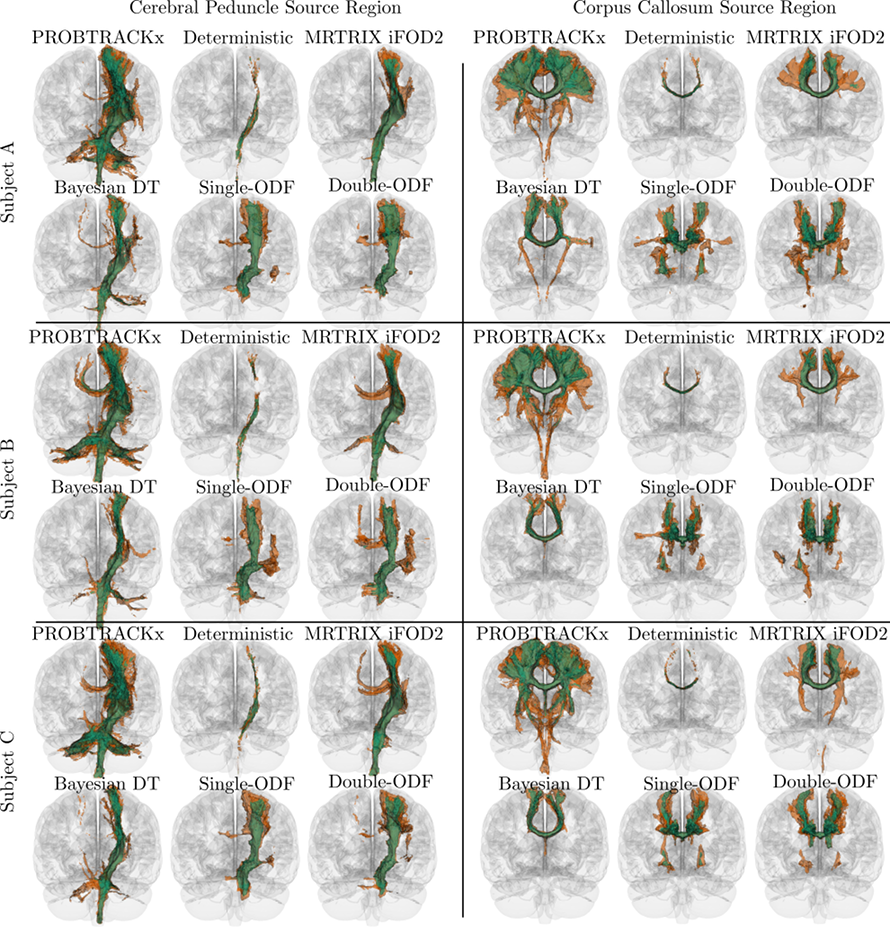}
    \caption{ Tractography results from three HCP subjects. Voxels 
        within the top \hl{ 5\% of connectivity scores are
	enclosed in an orange surface and the top 2\% are enclosed in a green surface}.
	 The tractography method used to generate the map is listed above
	the brain images. Images in the left panels are the result of tracking
	from the cerebral peduncle ROI, and the right result from tracking from the
	corpus callosum ROI.}
\label{fig:peduncle_comparison}
\end{figure}

\begin{figure}
\begin{minipage}{\textwidth}
\subsection*{Additional slice views and thresholds for tractography comparison}
\centering
\includegraphics[height=0.7\textheight]{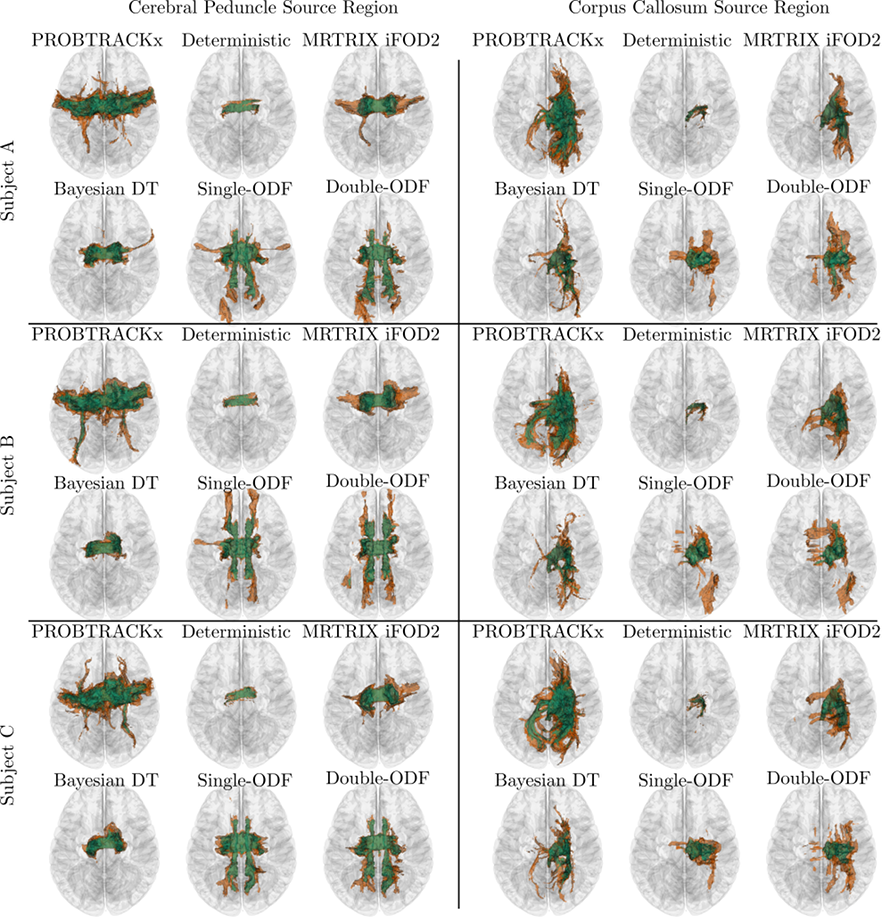}
\caption{ Tractography results from three HCP subjects. Voxels in the
    having the top \hl{ 5\% of connectivity scores are
	enclosed in an orange surface and the top 2\% are enclosed in a green surface}.
	The tractography method used to generate the map is listed above
	the brain images. Images in the left panels are the result of tracking
	from the cerebral peduncle ROI, and the right result from tracking from the
	corpus callosum ROI. \hl{This view is from the top of the brain facing downwards.}}
\label{fig:peduncle_comparison80}
\end{minipage}
\end{figure}

Following the approach of \citet{sotiropoulos2010brain} we extracted the
connectivity scores weights from all the nonzero-scored voxels output from each
method. In Figure~\ref{fig:peduncle_comparison} we plot 3D surfaces of the
voxels with scores in the top 5\% and 2\% of each method's results.  These are
displayed for the cerebral peduncle source region in
Figure~\ref{fig:peduncle_comparison} (left) and for the corpus callosum region
(right). These surfaces are displayed axially in
Figure~\ref{fig:peduncle_comparison80}.

Each method presented in Figure~\ref{fig:peduncle_comparison} behaves
consistently across all three subjects. All methods are able to find a path
from the peduncle to motor cortex, but the spatial extent of this connection
varies considerably.  Deterministic tractography is much more sparse than
the other methods, but this may be due to its conversion to a track density 
image (TDI). Most of the
million streamlines follow the same paths, which results in little variability
in the TDI, which in turn arbitrarily shrinks the spatial extent when
thresholding based on percentiles.

We compared spatial similarity of each method's output by calculating
Spearman's $\rho$ between the different methods\footnote{Spearman's
$\rho$ was chosen instead of Pearson's $r$ because the
connection weights estimated by probabilistic methods are approximately exponentially
distributed while analytic connection weights are \hl{approximately} normally distributed.
Therefore rank-based Spearman's $\rho$ was used.}.
These comparisons included all white matter voxels ($\approx900,000$) per subject.
\hl{Table~\ref{table:tracking_comparison} lists the within-subject $\rho$ values
across the three subjects.}  The double-ODF and iFOD2 results are an interesting
comparison since they both ``peek'' into neighboring ODFs at some point in
processing. Double-ODF incorporates the neighbor voxel's ODF directly into the
transition probability calculation while iFOD2 performs a weighted average of
the fODFs at each step of the probabilistic simulations. We observed that
iFOD2 produced the most similar results to analytic tractography with $\rho$ 
up to 0.71
for the corpus callosum ROI and up to 0.76 for the cerebral peduncle ROI.
In both cases these are the highest similarities observed across methodologies.

\begin{sidewaystable}
\centering
\caption{ {\Large Connectivity map similarity for 3 HCP subjects}}
\label{table:tracking_comparison}
Corpus Callosum
    \begin{tabular}{l*{6}{c}}
    Method      & Deterministic & MRTRIX iFOD2 &
                  Bayesian DT & Single-ODF  & Double-ODF \\
      \hline
      PROBTRACKx     &  27, 24, 23 &  62, 61, 63 &  49, 42, 45 &  51, 51, 53 &  51, 51, 53 \\
      Deterministic  &                &  24, 24, 21 &  28, 29, 3  &  16, 15, 16 &  15, 14, 16 \\
      MRTRIX iFOD2   &                &                &  37, 33, 34 &  64, 65, 71 &  64, 65, 71 \\
      Bayesian DT    &                &                &                &  26, 24, 27 &  27, 24, 28 \\
      Single-ODF     &                &                &                &                &  98, 98, 98 \\
    \end{tabular}
\\
Cerebral Peduncle
    \begin{tabular}{l*{6}{c}}
    Method      & Deterministic & MRTRIX iFOD2 &
                  Bayesian DT & Single-ODF  & Double-ODF \\
      \hline
      PROBTRACKx     & 28, 28, 28  & 62, 62, 63 & 56, 55, 53 & 48, 47, 47  & 49, 48, 49  \\
      Deterministic  &             & 28, 30, 28 & 31, 31, 33 & 22, 23, 21  & 23, 23, 22  \\
      MRTRIX iFOD2   &             &            & 32, 33, 32 & 77, 75, 74  & 78, 76, 76  \\
      Bayesian DT    &             &            &            & 22, 19, 21  & 22, 20, 21  \\
      Single-ODF     &             &            &            &             & 98, 98, 98  \\
    \end{tabular}

\bigskip
        \hl{
Spearman's $\rho$ calculated pairwise on connectivity maps from probabilistic, deterministic
        and analytic tractography methods. Values are presented are multiplied by 100 for easier reading.}
\end{sidewaystable}

\hl{There is a large difference between analytic tracking and the probabilistic
methods in computation time.} On average, probtrackx took 38 minutes per
region, Bayes Dirac took 38 minutes and iFOD took an estimated 5 hours.
Calculating a whole-brain shortest path map after calculating analytic
transition probabilities took an average of only 0.64 seconds. \hl{A single
query from a source region results in a probability value in each voxel, so
constructing a region-to-region connectivity matrix would take about 0.64
seconds per region in the atlas using voxel-grapn methods with analytic
transition probabilities as edge weights.}

\section{Discussion}

The analytic tractography framework provides fast, theoretically grounded tools
for quantifying white matter structure through dMRI data.  Closed-form
solutions for two types of transition probabilities provide precision that
simulation-based methods can only approximate. Moreover, by avoiding simulation,
the accuracy of analytic methods does not depend on the number of simulations run.
Furthermore, the analytic solutions can be calculated for any q-space sampling
scheme or ODF reconstruction method. This generality frees the experimenter
from an additional set of simulation-related parameters in an already complex
processing pipeline. The efficiency and precision of calculation enables
researchers to rapidly explore the effects of choices in other parameter spaces 
such as ODF reconstruction and deconvolution.

\hl{Although the low computational demands of analytic tractography are a desirable
feature, it can rightly be argued that the probabilistic simulations run by
most current software can also be run on GPUs at comparable speeds
\citep{hernandez2013accelerating}. We emphasize
that even though GPUs can simulate seeds quickly, they will still be simulating
a finite number of seeds in a subset of voxels -- leading to simulation-related
error. We have demonstrated that simulating $10^6$ seeds per voxel will produce
simulation-related errors on the order of $10^{-4}$, but this error rate only applies
to voxels explicitly included in a seed region. Voxels external to the seed region
will be sampled by exponentially fewer seeds and will have a corresponding
increase in error.}

\hl{We demonstrate empirically that the result of probabilistic tractography
converges to the analytic solution as the number of seeds increases. This
result should be viewed in the context of our comparison of the \iodfone method
to ground-truth transition probabilities.  Transition probabilities simulated
on one single, symmetric ODF at a time were markedly less accurate than
asymmetric transition probabilities calculated by the \iodftwo method that
incorporates multiple ODFs. This performance drop was observed in noise-free
simulated dMRI, meaning that more severe drops may be observed on noisy real
dMRI data. We see this as compelling motivation to adopt asymmetric ODF-capable
methods such those proposed by \citet{reisert2012geometry} and
\citet{bastiani2017improved}.}

The directed voxel graph built from transition probabilities opens up new
possibilities in brain network analysis. Current network science methods
primarily model data that are described at the anatomic resolution of large
cortical regions. They typically make use of tractography to weight edges.
These weights are sensitive to a large number of tractography reconstruction
parameters \citep{bassett2011conserved}.  In contrast, our voxel graph
application is insensitive to these reconstruction choices and will instead
efficiently evaluate all possible tractograms -- providing a probability for
each path. Furthermore, these path probabilities are free of simulation-based
errors that accumulate at each step in probabilistic tractography. The tracking
method presented here does have the previously noted distance bias
\citep{Iturria-Medina2011}, but this bias manifests as a regression-to-the-mean
instead of an exponential decay. Future work will investigate more robust
methods for calculating path probabilities.

Voxel graphs are computationally efficient. The graph only needs to be
constructed once for the whole brain, after which it can be queried from any
arbitrary location. Fortunately, finding the shortest, or even the top $k$
shortest paths in a graph is a computationally tractable problem. An arsenal of
efficient computational approaches do this well; a number of efficient
algorithms are known whose running time is essentially linear in the size of
the voxel graph \citep{dijkstra1959note}.  We observed up to a 2,812,500\%
improvement in processing time over comparable CPU-based probabilistic
tractography methods. Our method should accelerate the evolution of these
approaches.

\hl{We have not addressed the issue of how to construct region-to-region
structural connectivity matrices here.  There are already numerous methods
for using Voxel Graphs to estimate the connection strength between
regions. These include Anatomical Connection Strength,
Anatomical Connection Density, and Anatomical Connection Probability
as proposed by \citet{iturria2007characterizing} or capacity as
proposed by \citet{Zalesky2009}. \citet{sotiropoulos2010brain} also
propose a number of methods. Future work will investigate structural
connectivity matrices using Voxel Graphs.}

A unique avenue of connectivity quantification is possible with voxel
graphs.  Recent work shows that specific regions in deep white matter
are prone to a high rate of false connections in current tractography methods
\citep{maier2016tractography}. We could potentially identify these regions, as
we can now compute centrality measures in individual voxels.  In fact,
network measures that are commonly calculated for entire brain regions
can now be calculated voxelwise in all of white matter.  While we
demonstrate shortest path queries on voxel graphs, this is only one
example of the many opportunities to pursue novel connectivity metrics.
Our software is easy to install and use, providing a means for the
community to do so.

\hl{Future work should address the problem of large-scale geometric constraints
on shortest path trajectories.  Another direction for future research would be
an exhaustive test of the step size and turning angle max parameters over a
range of commonly-used voxel sizes. Using a more realistic fiber phantom such
as the one introduced by \citet{maier2016tractography} as ground-truth and
calculating error CDFs across a grid of step size and angle max parameters
would allow for fine-tuning these choices based on voxel sizes that could be
useful for other tractography methods as well. Once tuned, this method may
be useful for clinical applications that are typically approached with
tractography-based methods.

}

\section*{Acknowledgements}

This research was supported by a Head Health Challenge grant from
General-Electric and the National Football League and Institute for
Collaborative Biotechnologies through grant W911NF-09-0001 from the U.S.
Army Research Office.

\section*{References}
\bibliography{tn}

\begin{thebibliography}{45}
\expandafter\ifx\csname natexlab\endcsname\relax\def\natexlab#1{#1}\fi
\providecommand{\url}[1]{\texttt{#1}}
\providecommand{\href}[2]{#2}
\providecommand{\path}[1]{#1}
\providecommand{\DOIprefix}{doi:}
\providecommand{\ArXivprefix}{arXiv:}
\providecommand{\URLprefix}{URL: }
\providecommand{\Pubmedprefix}{pmid:}
\providecommand{\doi}[1]{\href{http://dx.doi.org/#1}{\path{#1}}}
\providecommand{\Pubmed}[1]{\href{pmid:#1}{\path{#1}}}
\providecommand{\bibinfo}[2]{#2}
\ifx\xfnm\relax \def\xfnm[#1]{\unskip,\space#1}\fi
\bibitem[{Avants et~al.(2007)Avants, Duda, Zhang \& Gee}]{Avants2007}
\bibinfo{author}{Avants, B.}, \bibinfo{author}{Duda, J.~T.},
  \bibinfo{author}{Zhang, H.}, \& \bibinfo{author}{Gee, J.~C.}
  (\bibinfo{year}{2007}).
\newblock \bibinfo{title}{{Multivariate Normalization with Symmetric
  Diffeomorphisms for Multivariate Studies}}.
\newblock In {\it \bibinfo{booktitle}{{Medical Image Computing and
  Computer-Assisted Intervention -- MICCAI 2007}}\/} (pp.
  \bibinfo{pages}{359--366}).
\newblock \bibinfo{address}{Berlin, Heidelberg}: \bibinfo{publisher}{Springer
  Berlin Heidelberg}.
\bibitem[{Avants et~al.(2008)Avants, Epstein, Grossman \&
  Gee}]{avants2008symmetric}
\bibinfo{author}{Avants, B.~B.}, \bibinfo{author}{Epstein, C.~L.},
  \bibinfo{author}{Grossman, M.}, \& \bibinfo{author}{Gee, J.~C.}
  (\bibinfo{year}{2008}).
\newblock \bibinfo{title}{{Symmetric diffeomorphic image registration with
  cross-correlation: evaluating automated labeling of elderly and
  neurodegenerative brain}}.
\newblock {\it \bibinfo{journal}{Medical image analysis}\/},  {\it
  \bibinfo{volume}{12}\/}, \bibinfo{pages}{26--41}.
\bibitem[{Avants et~al.(2010)Avants, Yushkevich, Pluta, Minkoff, Korczykowski,
  Detre \& Gee}]{avants2010optimal}
\bibinfo{author}{Avants, B.~B.}, \bibinfo{author}{Yushkevich, P.},
  \bibinfo{author}{Pluta, J.}, \bibinfo{author}{Minkoff, D.},
  \bibinfo{author}{Korczykowski, M.}, \bibinfo{author}{Detre, J.}, \&
  \bibinfo{author}{Gee, J.~C.} (\bibinfo{year}{2010}).
\newblock \bibinfo{title}{{The optimal template effect in hippocampus studies
  of diseased populations}}.
\newblock {\it \bibinfo{journal}{Neuroimage}\/},  {\it \bibinfo{volume}{49}\/},
  \bibinfo{pages}{2457--2466}.
\bibitem[{Basser et~al.(2000)Basser, Pajevic, Pierpaoli, Duda \&
  Aldroubi}]{Basser2000}
\bibinfo{author}{Basser, P.}, \bibinfo{author}{Pajevic, S.},
  \bibinfo{author}{Pierpaoli, C.}, \bibinfo{author}{Duda, J.}, \&
  \bibinfo{author}{Aldroubi, A.} (\bibinfo{year}{2000}).
\newblock \bibinfo{title}{{In vivo fiber tractography using DT-MRI data}}.
\newblock {\it \bibinfo{journal}{Magnetic Resonance in Medicine}\/},  {\it
  \bibinfo{volume}{44}\/}, \bibinfo{pages}{625--632}.
\bibitem[{Bassett et~al.(2011)Bassett, Brown, Deshpande, Carlson \&
  Grafton}]{bassett2011conserved}
\bibinfo{author}{Bassett, D.~S.}, \bibinfo{author}{Brown, J.~A.},
  \bibinfo{author}{Deshpande, V.}, \bibinfo{author}{Carlson, J.~M.}, \&
  \bibinfo{author}{Grafton, S.~T.} (\bibinfo{year}{2011}).
\newblock \bibinfo{title}{{Conserved and variable architecture of human white
  matter connectivity}}.
\newblock {\it \bibinfo{journal}{Neuroimage}\/},  {\it \bibinfo{volume}{54}\/},
  \bibinfo{pages}{1262--1279}.
\bibitem[{Bastiani et~al.(2017)Bastiani, Cottaar, Dikranian, Ghosh, Zhang,
  Alexander, Behrens, Jbabdi \& Sotiropoulos}]{bastiani2017improved}
\bibinfo{author}{Bastiani, M.}, \bibinfo{author}{Cottaar, M.},
  \bibinfo{author}{Dikranian, K.}, \bibinfo{author}{Ghosh, A.},
  \bibinfo{author}{Zhang, H.}, \bibinfo{author}{Alexander, D.~C.},
  \bibinfo{author}{Behrens, T.~E.}, \bibinfo{author}{Jbabdi, S.}, \&
  \bibinfo{author}{Sotiropoulos, S.~N.} (\bibinfo{year}{2017}).
\newblock \bibinfo{title}{Improved tractography using asymmetric fibre
  orientation distributions}.
\newblock {\it \bibinfo{journal}{NeuroImage}\/},  {\it
  \bibinfo{volume}{158}\/}, \bibinfo{pages}{205--218}.
\bibitem[{Behrens et~al.(2007)Behrens, Berg, Jbabdi, Rushworth \&
  Woolrich}]{behrens2007probabilistic}
\bibinfo{author}{Behrens, T.~E.}, \bibinfo{author}{Berg, H.~J.},
  \bibinfo{author}{Jbabdi, S.}, \bibinfo{author}{Rushworth, M.}, \&
  \bibinfo{author}{Woolrich, M.} (\bibinfo{year}{2007}).
\newblock \bibinfo{title}{{Probabilistic diffusion tractography with multiple
  fibre orientations: What can we gain?}}
\newblock {\it \bibinfo{journal}{Neuroimage}\/},  {\it \bibinfo{volume}{34}\/},
  \bibinfo{pages}{144--155}.
\bibitem[{Behrens et~al.(2003)Behrens, Woolrich, Jenkinson, Johansen-Berg,
  Nunes, Clare, Matthews, Brady \& Smith}]{behrens2003characterization}
\bibinfo{author}{Behrens, T.~E.}, \bibinfo{author}{Woolrich, M.},
  \bibinfo{author}{Jenkinson, M.}, \bibinfo{author}{Johansen-Berg, H.},
  \bibinfo{author}{Nunes, R.}, \bibinfo{author}{Clare, S.},
  \bibinfo{author}{Matthews, P.}, \bibinfo{author}{Brady, J.~M.}, \&
  \bibinfo{author}{Smith, S.~M.} (\bibinfo{year}{2003}).
\newblock \bibinfo{title}{{Characterization and propagation of uncertainty in
  diffusion-weighted MR imaging}}.
\newblock {\it \bibinfo{journal}{Magnetic resonance in medicine}\/},  {\it
  \bibinfo{volume}{50}\/}, \bibinfo{pages}{1077--1088}.
\bibitem[{Calamante et~al.(2010)Calamante, Tournier, Jackson \&
  Connelly}]{Calamante2010}
\bibinfo{author}{Calamante, F.}, \bibinfo{author}{Tournier, J.~D.},
  \bibinfo{author}{Jackson, G.~D.}, \& \bibinfo{author}{Connelly, A.}
  (\bibinfo{year}{2010}).
\newblock \bibinfo{title}{{Track-density imaging (TDI): Super-resolution white
  matter imaging using whole-brain track-density mapping}}.
\newblock {\it \bibinfo{journal}{NeuroImage}\/},  {\it \bibinfo{volume}{53}\/},
  \bibinfo{pages}{1233--1243}.
\bibitem[{Campbell et~al.(2005)Campbell, Siddiqi, Rymar, Sadikot \&
  Pike}]{Campbell2005}
\bibinfo{author}{Campbell, J. S.~W.}, \bibinfo{author}{Siddiqi, K.},
  \bibinfo{author}{Rymar, V.~V.}, \bibinfo{author}{Sadikot, A.~F.}, \&
  \bibinfo{author}{Pike, G.~B.} (\bibinfo{year}{2005}).
\newblock \bibinfo{title}{{Flow-based fiber tracking with diffusion tensor and
  q-ball data: Validation and comparison to principal diffusion direction
  techniques}}.
\newblock {\it \bibinfo{journal}{NeuroImage}\/},  {\it \bibinfo{volume}{27}\/},
  \bibinfo{pages}{725--736}. \DOIprefix\doi{10.1016/j.neuroimage.2005.05.014}.
\bibitem[{Descoteaux et~al.(2009)Descoteaux, Deriche, Knosche \&
  Anwander}]{descoteaux2009deterministic}
\bibinfo{author}{Descoteaux, M.}, \bibinfo{author}{Deriche, R.},
  \bibinfo{author}{Knosche, T.~R.}, \& \bibinfo{author}{Anwander, A.}
  (\bibinfo{year}{2009}).
\newblock \bibinfo{title}{{Deterministic and probabilistic tractography based
  on complex fibre orientation distributions}}.
\newblock {\it \bibinfo{journal}{IEEE transactions on medical imaging}\/},
  {\it \bibinfo{volume}{28}\/}, \bibinfo{pages}{269--286}.
\bibitem[{Dhollander et~al.(2014)Dhollander, Emsell, Van~Hecke, Maes, Sunaert
  \& Suetens}]{dhollander2014track}
\bibinfo{author}{Dhollander, T.}, \bibinfo{author}{Emsell, L.},
  \bibinfo{author}{Van~Hecke, W.}, \bibinfo{author}{Maes, F.},
  \bibinfo{author}{Sunaert, S.}, \& \bibinfo{author}{Suetens, P.}
  (\bibinfo{year}{2014}).
\newblock \bibinfo{title}{Track orientation density imaging (todi) and track
  orientation distribution (tod) based tractography}.
\newblock {\it \bibinfo{journal}{NeuroImage}\/},  {\it \bibinfo{volume}{94}\/},
  \bibinfo{pages}{312--336}.
\bibitem[{Dijkstra(1959)}]{dijkstra1959note}
\bibinfo{author}{Dijkstra, E.~W.} (\bibinfo{year}{1959}).
\newblock \bibinfo{title}{{A note on two problems in connexion with graphs}}.
\newblock {\it \bibinfo{journal}{Numerische mathematik}\/},  {\it
  \bibinfo{volume}{1}\/}, \bibinfo{pages}{269--271}.
\bibitem[{Fillard et~al.(2011)Fillard, Descoteaux, Goh, Gouttard, Jeurissen,
  Malcolm, Ramirez-Manzanares, Reisert, Sakaie, Tensaouti
  et~al.}]{fillard2011quantitative}
\bibinfo{author}{Fillard, P.}, \bibinfo{author}{Descoteaux, M.},
  \bibinfo{author}{Goh, A.}, \bibinfo{author}{Gouttard, S.},
  \bibinfo{author}{Jeurissen, B.}, \bibinfo{author}{Malcolm, J.},
  \bibinfo{author}{Ramirez-Manzanares, A.}, \bibinfo{author}{Reisert, M.},
  \bibinfo{author}{Sakaie, K.}, \bibinfo{author}{Tensaouti, F.} et~al.
  (\bibinfo{year}{2011}).
\newblock \bibinfo{title}{Quantitative evaluation of 10 tractography algorithms
  on a realistic diffusion mr phantom}.
\newblock {\it \bibinfo{journal}{Neuroimage}\/},  {\it \bibinfo{volume}{56}\/},
  \bibinfo{pages}{220--234}.
\bibitem[{Friman et~al.(2006)Friman, Farneback \& Westin}]{friman2006bayesian}
\bibinfo{author}{Friman, O.}, \bibinfo{author}{Farneback, G.}, \&
  \bibinfo{author}{Westin, C.-F.} (\bibinfo{year}{2006}).
\newblock \bibinfo{title}{{A Bayesian approach for stochastic white matter
  tractography}}.
\newblock {\it \bibinfo{journal}{IEEE transactions on medical imaging}\/},
  {\it \bibinfo{volume}{25}\/}, \bibinfo{pages}{965--978}.
\bibitem[{Garyfallidis et~al.(2014)Garyfallidis, Brett, Amirbekian, Rokem,
  van~der Walt, Descoteaux \& Nimmo-Smith}]{Garyfallidis2014}
\bibinfo{author}{Garyfallidis, E.}, \bibinfo{author}{Brett, M.},
  \bibinfo{author}{Amirbekian, B.}, \bibinfo{author}{Rokem, A.},
  \bibinfo{author}{van~der Walt, S.}, \bibinfo{author}{Descoteaux, M.}, \&
  \bibinfo{author}{Nimmo-Smith, I.} (\bibinfo{year}{2014}).
\newblock \bibinfo{title}{{Dipy, a library for the analysis of diffusion MRI
  data.}}
\newblock {\it \bibinfo{journal}{Frontiers in neuroinformatics}\/},  {\it
  \bibinfo{volume}{8}\/}, \bibinfo{pages}{8}.
\bibitem[{Glasser et~al.(2013)Glasser, Sotiropoulos, Wilson, Coalson, Fischl,
  Andersson, Xu, Jbabdi, Webster, Polimeni et~al.}]{glasser2013minimal}
\bibinfo{author}{Glasser, M.~F.}, \bibinfo{author}{Sotiropoulos, S.~N.},
  \bibinfo{author}{Wilson, J.~A.}, \bibinfo{author}{Coalson, T.~S.},
  \bibinfo{author}{Fischl, B.}, \bibinfo{author}{Andersson, J.~L.},
  \bibinfo{author}{Xu, J.}, \bibinfo{author}{Jbabdi, S.},
  \bibinfo{author}{Webster, M.}, \bibinfo{author}{Polimeni, J.~R.} et~al.
  (\bibinfo{year}{2013}).
\newblock \bibinfo{title}{{The minimal preprocessing pipelines for the Human
  Connectome Project}}.
\newblock {\it \bibinfo{journal}{Neuroimage}\/},  {\it \bibinfo{volume}{80}\/},
  \bibinfo{pages}{105--124}.
\bibitem[{Hageman et~al.(2009)Hageman, Toga, Narr \& Shattuck}]{Hageman2009}
\bibinfo{author}{Hageman, N.}, \bibinfo{author}{Toga, A.},
  \bibinfo{author}{Narr, K.}, \& \bibinfo{author}{Shattuck, D.}
  (\bibinfo{year}{2009}).
\newblock \bibinfo{title}{{A diffusion tensor imaging tractography algorithm
  based on Navier--Stokes fluid mechanics}}.
\newblock {\it \bibinfo{journal}{Medical Imaging, IEEE Transactions on}\/},
  {\it \bibinfo{volume}{28}\/}, \bibinfo{pages}{348--360}.
\bibitem[{Hern{\'a}ndez et~al.(2013)Hern{\'a}ndez, Guerrero, Cecilia,
  Garc{\'i}a, Inuggi, Jbabdi, Behrens \&
  Sotiropoulos}]{hernandez2013accelerating}
\bibinfo{author}{Hern{\'a}ndez, M.}, \bibinfo{author}{Guerrero, G.~D.},
  \bibinfo{author}{Cecilia, J.~M.}, \bibinfo{author}{Garc{\'i}a, J.~M.},
  \bibinfo{author}{Inuggi, A.}, \bibinfo{author}{Jbabdi, S.},
  \bibinfo{author}{Behrens, T.~E.}, \& \bibinfo{author}{Sotiropoulos, S.~N.}
  (\bibinfo{year}{2013}).
\newblock \bibinfo{title}{{Accelerating fibre orientation estimation from
  diffusion weighted magnetic resonance imaging using GPUs}}.
\newblock {\it \bibinfo{journal}{PLoS One}\/},  {\it \bibinfo{volume}{8}\/},
  \bibinfo{pages}{e61892}.
\bibitem[{Iturria-Medina et~al.(2007)Iturria-Medina, Canales-Rodriguez,
  Melie-Garcia, Valdes-Hernandez, Martinez-Montes, Alem{\'a}n-G{\'o}mez \&
  S{\'a}nchez-Bornot}]{iturria2007characterizing}
\bibinfo{author}{Iturria-Medina, Y.}, \bibinfo{author}{Canales-Rodriguez, E.},
  \bibinfo{author}{Melie-Garcia, L.}, \bibinfo{author}{Valdes-Hernandez, P.},
  \bibinfo{author}{Martinez-Montes, E.}, \bibinfo{author}{Alem{\'a}n-G{\'o}mez,
  Y.}, \& \bibinfo{author}{S{\'a}nchez-Bornot, J.} (\bibinfo{year}{2007}).
\newblock \bibinfo{title}{{Characterizing brain anatomical connections using
  diffusion weighted MRI and graph theory}}.
\newblock {\it \bibinfo{journal}{Neuroimage}\/},  {\it \bibinfo{volume}{36}\/},
  \bibinfo{pages}{645--660}.
\bibitem[{Iturria-Medina et~al.(2011)Iturria-Medina, {P{\'e}rez Fern{\'a}ndez},
  Morris, Canales-Rodr{\'i}guez, Haroon, {Garc{\'i}a Pent{\'o}n}, Augath,
  {Gal{\'a}n Garc{\'i}a}, Logothetis, Parker \&
  Melie-Garc{\'i}a}]{Iturria-Medina2011}
\bibinfo{author}{Iturria-Medina, Y.}, \bibinfo{author}{{P{\'e}rez
  Fern{\'a}ndez}, A.}, \bibinfo{author}{Morris, D.~M.},
  \bibinfo{author}{Canales-Rodr{\'i}guez, E.~J.}, \bibinfo{author}{Haroon,
  H.~A.}, \bibinfo{author}{{Garc{\'i}a Pent{\'o}n}, L.},
  \bibinfo{author}{Augath, M.}, \bibinfo{author}{{Gal{\'a}n Garc{\'i}a}, L.},
  \bibinfo{author}{Logothetis, N.}, \bibinfo{author}{Parker, G. J.~M.}, \&
  \bibinfo{author}{Melie-Garc{\'i}a, L.} (\bibinfo{year}{2011}).
\newblock \bibinfo{title}{{Brain hemispheric structural efficiency and
  interconnectivity rightward asymmetry in human and nonhuman primates.}}
\newblock {\it \bibinfo{journal}{Cerebral cortex (New York, N.Y. : 1991)}\/},
  {\it \bibinfo{volume}{21}\/}, \bibinfo{pages}{56--67}.
\bibitem[{Iturria-Medina et~al.(2008)Iturria-Medina, Sotero,
  Canales-Rodr{\'\i}guez, Alem{\'a}n-G{\'o}mez \&
  Melie-Garc{\'\i}a}]{iturria2008studying}
\bibinfo{author}{Iturria-Medina, Y.}, \bibinfo{author}{Sotero, R.~C.},
  \bibinfo{author}{Canales-Rodr{\'\i}guez, E.~J.},
  \bibinfo{author}{Alem{\'a}n-G{\'o}mez, Y.}, \&
  \bibinfo{author}{Melie-Garc{\'\i}a, L.} (\bibinfo{year}{2008}).
\newblock \bibinfo{title}{Studying the human brain anatomical network via
  diffusion-weighted mri and graph theory}.
\newblock {\it \bibinfo{journal}{Neuroimage}\/},  {\it \bibinfo{volume}{40}\/},
  \bibinfo{pages}{1064--1076}.
\bibitem[{Jeurissen et~al.(2014)Jeurissen, Tournier, Dhollander, Connelly \&
  Sijbers}]{jeurissen2014multi}
\bibinfo{author}{Jeurissen, B.}, \bibinfo{author}{Tournier, J.-D.},
  \bibinfo{author}{Dhollander, T.}, \bibinfo{author}{Connelly, A.}, \&
  \bibinfo{author}{Sijbers, J.} (\bibinfo{year}{2014}).
\newblock \bibinfo{title}{{Multi-tissue constrained spherical deconvolution for
  improved analysis of multi-shell diffusion MRI data}}.
\newblock {\it \bibinfo{journal}{NeuroImage}\/},  {\it
  \bibinfo{volume}{103}\/}, \bibinfo{pages}{411--426}.
\bibitem[{Jones et~al.(2013)Jones, Kn{\"o}sche \& Turner}]{Jones2013}
\bibinfo{author}{Jones, D.~K.}, \bibinfo{author}{Kn{\"o}sche, T.~R.}, \&
  \bibinfo{author}{Turner, R.} (\bibinfo{year}{2013}).
\newblock \bibinfo{title}{{White matter integrity, fiber count, and other
  fallacies: The do's and don'ts of diffusion MRI}}.
\newblock {\it \bibinfo{journal}{NeuroImage}\/},  {\it \bibinfo{volume}{73}\/},
  \bibinfo{pages}{239--254}.
\bibitem[{Kaden et~al.(2007)Kaden, Kn{\"o}sche \&
  Anwander}]{kaden2007parametric}
\bibinfo{author}{Kaden, E.}, \bibinfo{author}{Kn{\"o}sche, T.~R.}, \&
  \bibinfo{author}{Anwander, A.} (\bibinfo{year}{2007}).
\newblock \bibinfo{title}{{Parametric spherical deconvolution: Inferring
  anatomical connectivity using diffusion MR imaging}}.
\newblock {\it \bibinfo{journal}{NeuroImage}\/},  {\it \bibinfo{volume}{37}\/},
  \bibinfo{pages}{474--488}.
\bibitem[{Koch et~al.(2016)Koch, Staudt, Vogel \&
  Meyerhenke}]{koch2016empirical}
\bibinfo{author}{Koch, J.}, \bibinfo{author}{Staudt, C.~L.},
  \bibinfo{author}{Vogel, M.}, \& \bibinfo{author}{Meyerhenke, H.}
  (\bibinfo{year}{2016}).
\newblock \bibinfo{title}{{An empirical comparison of Big Graph frameworks in
  the context of network analysis}}.
\newblock {\it \bibinfo{journal}{Social Network Analysis and Mining}\/},  {\it
  \bibinfo{volume}{6}\/}, \bibinfo{pages}{84}.
\bibitem[{Koch et~al.(2002)Koch, Norris \&
  Hund-Georgiadis}]{koch2002investigation}
\bibinfo{author}{Koch, M.~A.}, \bibinfo{author}{Norris, D.~G.}, \&
  \bibinfo{author}{Hund-Georgiadis, M.} (\bibinfo{year}{2002}).
\newblock \bibinfo{title}{{An Investigation of Functional and Anatomical
  Connectivity Using Magnetic Resonance Imaging}}.
\newblock {\it \bibinfo{journal}{NeuroImage}\/},  {\it \bibinfo{volume}{16}\/},
  \bibinfo{pages}{241--250}. \DOIprefix\doi{10.1006/nimg.2001.1052}.
\bibitem[{Maier-Hein et~al.(2016)Maier-Hein, Neher, Houde, Cote, Garyfallidis,
  Zhong, Chamberland, Yeh, Lin, Ji et~al.}]{maier2016tractography}
\bibinfo{author}{Maier-Hein, K.}, \bibinfo{author}{Neher, P.},
  \bibinfo{author}{Houde, J.-C.}, \bibinfo{author}{Cote, M.-A.},
  \bibinfo{author}{Garyfallidis, E.}, \bibinfo{author}{Zhong, J.},
  \bibinfo{author}{Chamberland, M.}, \bibinfo{author}{Yeh, F.-C.},
  \bibinfo{author}{Lin, Y.~C.}, \bibinfo{author}{Ji, Q.} et~al.
  (\bibinfo{year}{2016}).
\newblock \bibinfo{title}{{Tractography-based connectomes are dominated by
  false-positive connections}}.
\newblock {\it \bibinfo{journal}{bioRxiv}\/},  (p. \bibinfo{pages}{084137}).
\bibitem[{Neher et~al.(2014)Neher, Laun, Stieltjes \&
  Maier-Hein}]{neher2014fiberfox}
\bibinfo{author}{Neher, P.~F.}, \bibinfo{author}{Laun, F.~B.},
  \bibinfo{author}{Stieltjes, B.}, \& \bibinfo{author}{Maier-Hein, K.~H.}
  (\bibinfo{year}{2014}).
\newblock \bibinfo{title}{Fiberfox: facilitating the creation of realistic
  white matter software phantoms}.
\newblock {\it \bibinfo{journal}{Magnetic resonance in medicine}\/},  {\it
  \bibinfo{volume}{72}\/}, \bibinfo{pages}{1460--1470}.
\bibitem[{Parker \& Alexander(2003)}]{parker2003probabilistic}
\bibinfo{author}{Parker, G.~J.}, \& \bibinfo{author}{Alexander, D.~C.}
  (\bibinfo{year}{2003}).
\newblock \bibinfo{title}{{Probabilistic Monte Carlo based mapping of cerebral
  connections utilising whole-brain crossing fibre information}}.
\newblock In {\it \bibinfo{booktitle}{{Biennial International Conference on
  Information Processing in Medical Imaging}}\/} (pp.
  \bibinfo{pages}{684--695}).
\newblock \bibinfo{organization}{Springer}.
\bibitem[{Peterson(2009)}]{peterson2009f2py}
\bibinfo{author}{Peterson, P.} (\bibinfo{year}{2009}).
\newblock \bibinfo{title}{{F2PY: a tool for connecting Fortran and Python
  programs}}.
\newblock {\it \bibinfo{journal}{International Journal of Computational Science
  and Engineering}\/},  {\it \bibinfo{volume}{4}\/}, \bibinfo{pages}{296--305}.
\bibitem[{Poupon et~al.(2010)Poupon, Laribiere, Tournier, Bernard, Fournier,
  Fillard, Descoteaux \& Mangin}]{poupon2010diffusion}
\bibinfo{author}{Poupon, C.}, \bibinfo{author}{Laribiere, L.},
  \bibinfo{author}{Tournier, G.}, \bibinfo{author}{Bernard, J.},
  \bibinfo{author}{Fournier, D.}, \bibinfo{author}{Fillard, P.},
  \bibinfo{author}{Descoteaux, M.}, \& \bibinfo{author}{Mangin, J.}
  (\bibinfo{year}{2010}).
\newblock \bibinfo{title}{A diffusion hardware phantom looking like a coronal
  brain slice}.
\newblock In {\it \bibinfo{booktitle}{Proceedings of the International Society
  for Magnetic Resonance in Medicine}\/} (p. \bibinfo{pages}{581}).
\newblock volume~\bibinfo{volume}{18}.
\bibitem[{Reisert et~al.(2012)Reisert, Kellner \&
  Kiselev}]{reisert2012geometry}
\bibinfo{author}{Reisert, M.}, \bibinfo{author}{Kellner, E.}, \&
  \bibinfo{author}{Kiselev, V.~G.} (\bibinfo{year}{2012}).
\newblock \bibinfo{title}{About the geometry of asymmetric fiber orientation
  distributions}.
\newblock {\it \bibinfo{journal}{IEEE transactions on medical imaging}\/},
  {\it \bibinfo{volume}{31}\/}, \bibinfo{pages}{1240--1249}.
\bibitem[{Sotiropoulos et~al.(2010)Sotiropoulos, Bai, Morgan, Constantinescu \&
  Tench}]{sotiropoulos2010brain}
\bibinfo{author}{Sotiropoulos, S.~N.}, \bibinfo{author}{Bai, L.},
  \bibinfo{author}{Morgan, P.~S.}, \bibinfo{author}{Constantinescu, C.~S.}, \&
  \bibinfo{author}{Tench, C.~R.} (\bibinfo{year}{2010}).
\newblock \bibinfo{title}{{Brain tractography using Q-ball imaging and graph
  theory: Improved connectivities through fibre crossings via a model-based
  approach}}.
\newblock {\it \bibinfo{journal}{NeuroImage}\/},  {\it \bibinfo{volume}{49}\/},
  \bibinfo{pages}{2444--2456}.
\bibitem[{Staudt et~al.(2014)Staudt, Sazonovs \&
  Meyerhenke}]{staudt2014networkit}
\bibinfo{author}{Staudt, C.}, \bibinfo{author}{Sazonovs, A.}, \&
  \bibinfo{author}{Meyerhenke, H.} (\bibinfo{year}{2014}).
\newblock \bibinfo{title}{{Networkit: An interactive tool suite for
  high-performance network analysis}}.
\newblock {\it \bibinfo{journal}{CoRR, abs/1403.3005}\/}, .
\bibitem[{Tournier et~al.(2007)Tournier, Calamante \& Connelly}]{Tournier2007}
\bibinfo{author}{Tournier, J.~D.}, \bibinfo{author}{Calamante, F.}, \&
  \bibinfo{author}{Connelly, A.} (\bibinfo{year}{2007}).
\newblock \bibinfo{title}{{Robust determination of the fibre orientation
  distribution in diffusion MRI: Non-negativity constrained super-resolved
  spherical deconvolution}}.
\newblock {\it \bibinfo{journal}{NeuroImage}\/},  {\it \bibinfo{volume}{35}\/},
  \bibinfo{pages}{1459--1472}.
\bibitem[{Tournier et~al.(2010)Tournier, Calamante \& Connelly}]{Tournier2010}
\bibinfo{author}{Tournier, J.~D.}, \bibinfo{author}{Calamante, F.}, \&
  \bibinfo{author}{Connelly, A.} (\bibinfo{year}{2010}).
\newblock \bibinfo{title}{{Improved probabilistic streamlines tractography by
  2nd order integration over fibre orientation distributions}}.
\newblock {\it \bibinfo{journal}{Proceedings of the International Society for
  Magnetic Resonance in Medicine}\/},  (p. \bibinfo{pages}{1670}).
\bibitem[{Tournier et~al.(2012)Tournier, Calamante \& Connelly}]{Tournier2012}
\bibinfo{author}{Tournier, J.~D.}, \bibinfo{author}{Calamante, F.}, \&
  \bibinfo{author}{Connelly, A.} (\bibinfo{year}{2012}).
\newblock \bibinfo{title}{{MRtrix: Diffusion tractography in crossing fiber
  regions}}.
\newblock {\it \bibinfo{journal}{International Journal of Imaging Systems and
  Technology}\/},  {\it \bibinfo{volume}{22}\/}, \bibinfo{pages}{53--66}.
  \DOIprefix\doi{10.1002/ima.22005}.
\bibitem[{{Van Essen} \& Ugurbil(2012)}]{VanEssen2012}
\bibinfo{author}{{Van Essen}, D.~C.}, \& \bibinfo{author}{Ugurbil, K.}
  (\bibinfo{year}{2012}).
\newblock \bibinfo{title}{{The future of the human connectome}}.
\newblock {\it \bibinfo{journal}{NeuroImage}\/},  {\it \bibinfo{volume}{62}\/},
  \bibinfo{pages}{1299--1310}.
\bibitem[{Wedeen et~al.(2005)Wedeen, Hagmann, Tseng, Reese \&
  Weisskoff}]{Wedeen2005}
\bibinfo{author}{Wedeen, V.~J.}, \bibinfo{author}{Hagmann, P.},
  \bibinfo{author}{Tseng, W.-Y.~I.}, \bibinfo{author}{Reese, T.~G.}, \&
  \bibinfo{author}{Weisskoff, R.~M.} (\bibinfo{year}{2005}).
\newblock \bibinfo{title}{{Mapping complex tissue architecture with diffusion
  spectrum magnetic resonance imaging}}.
\newblock {\it \bibinfo{journal}{Magnetic Resonance in Medicine}\/},  {\it
  \bibinfo{volume}{54}\/}, \bibinfo{pages}{1377--1386}.
\bibitem[{Wedeen et~al.(2008)Wedeen, Wang, Schmahmann, Benner, Tseng, Dai,
  Pandya, Hagmann, D'Arceuil \& de~Crespigny}]{Wedeen2008}
\bibinfo{author}{Wedeen, V.~J.}, \bibinfo{author}{Wang, R.~P.},
  \bibinfo{author}{Schmahmann, J.~D.}, \bibinfo{author}{Benner, T.},
  \bibinfo{author}{Tseng, W. Y.~I.}, \bibinfo{author}{Dai, G.},
  \bibinfo{author}{Pandya, D.~N.}, \bibinfo{author}{Hagmann, P.},
  \bibinfo{author}{D'Arceuil, H.}, \& \bibinfo{author}{de~Crespigny, A.~J.}
  (\bibinfo{year}{2008}).
\newblock \bibinfo{title}{{Diffusion spectrum magnetic resonance imaging (DSI)
  tractography of crossing fibers}}.
\newblock {\it \bibinfo{journal}{NeuroImage}\/},  {\it \bibinfo{volume}{41}\/},
  \bibinfo{pages}{1267--1277}.
\bibitem[{Yeh et~al.(2013)Yeh, Verstynen, Wang, Fern{\'a}ndez-Miranda \&
  Tseng}]{Yeh2013deterministic}
\bibinfo{author}{Yeh, F.-C.}, \bibinfo{author}{Verstynen, T.~D.},
  \bibinfo{author}{Wang, Y.}, \bibinfo{author}{Fern{\'a}ndez-Miranda, J.~C.},
  \& \bibinfo{author}{Tseng, W.-Y.~I.} (\bibinfo{year}{2013}).
\newblock \bibinfo{title}{Deterministic diffusion fiber tracking improved by
  quantitative anisotropy}.
\newblock {\it \bibinfo{journal}{PloS one}\/},  {\it \bibinfo{volume}{8}\/},
  \bibinfo{pages}{e80713}.
\bibitem[{Yeh et~al.(2010)Yeh, Wedeen \& Tseng}]{Yeh2010}
\bibinfo{author}{Yeh, F.-C.}, \bibinfo{author}{Wedeen, V.~J.}, \&
  \bibinfo{author}{Tseng, W.-Y.~I.} (\bibinfo{year}{2010}).
\newblock \bibinfo{title}{{Generalized q-Sampling Imaging}}.
\newblock {\it \bibinfo{journal}{IEEE Transactions on Medical Imaging}\/},
  {\it \bibinfo{volume}{29}\/}, \bibinfo{pages}{1626--1635}.
\bibitem[{Zalesky(2008)}]{zalesky2008dt}
\bibinfo{author}{Zalesky, A.} (\bibinfo{year}{2008}).
\newblock \bibinfo{title}{{DT-MRI fiber tracking: a shortest paths approach}}.
\newblock {\it \bibinfo{journal}{IEEE transactions on medical imaging}\/},
  {\it \bibinfo{volume}{27}\/}, \bibinfo{pages}{1458--1471}.
\bibitem[{Zalesky \& Fornito(2009)}]{Zalesky2009}
\bibinfo{author}{Zalesky, A.}, \& \bibinfo{author}{Fornito, A.}
  (\bibinfo{year}{2009}).
\newblock \bibinfo{title}{{A DTI-Derived Measure of Cortico-Cortical
  Connectivity}}, .
\newblock (pp. \bibinfo{pages}{1--15}).

\end{thebibliography}

\newpage
\beginsupplement

\section*{Supplementary Materials}

\subsection{Implementation details}

We implemented analytic tractography as a Python (3.3+) package.  {\tt MITTENS}
functionality can be split into two areas. Equations~\ref{eq:transition_prob}
and \ref{eq:iodftwo} are implemented as Fortran90 extension modules
\citep{peterson2009f2py}. By default {\tt MITTENS} includes Fortran code to
calculate transition probabilities for the maximum turning angle and step size
used throughout this manuscript. {\tt MITTENS} can easily generate extension
modules for other parameter choices.

{\tt MITTENS} can read output from DSI Studio, which can reconstruct DTI, HARDI
and DSI and apply a number of deconvolution algorithms. We recommend
reconstructing and inspecting data for quality in DSI Studio before analysis
with {\tt MITTENS}.  Once a satisfactory reconstruction is saved {\tt MITTENS}
reads in the {\tt fib.gz} file and calculates transition probabilities from all
available voxels and their neighbors. \hl{The {\tt fib.gz} format is a MATLAB
file with a specific set of variables describing the spatial extent of the data,
voxel-size, anisotropy scalars and ODF data sampled on a standardized icosahedron.
Effort is currently underway to also load results from reconstructing in DiPy 
\citep{Garyfallidis2014} and MRTRIX \citep{Tournier2010}.
Transition probabilities} can be saved as a sparse matrix in
MATLAB format or a series of NIfTI-1 image files.  Calculating single-ODF and
double-ODF transition probabilities then takes approximately 27 minutes on a single
cpu. This calculation only needs to be performed once per reconstructed
dataset, after which the probabilities can be loaded directly into a graph
object. Our software uses {\tt networkit} \citep{staudt2014networkit}, a top
performing scalable graph library \citep{koch2016empirical}, to store and
search through the voxel graph.

\subsection*{Algorithm stability}

\begin{figure}[!hb]
\centering
\includegraphics[width=0.8\textwidth]{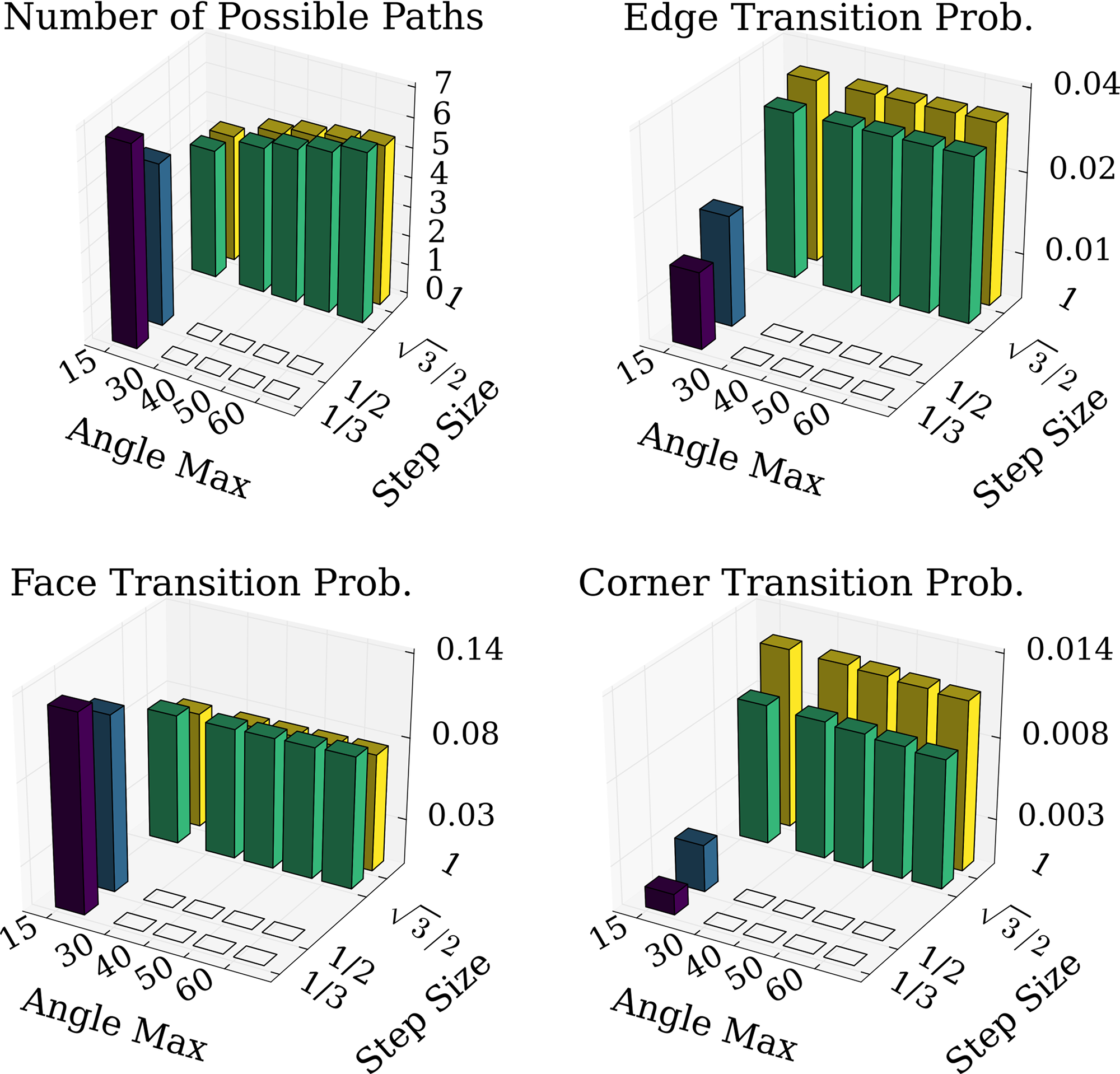}
\caption{Transition probabilities for an isotropic ODF are plotted
	as a function of step size and maximum turning angle. The $z$ axis of the
	top left plot is in $\log_{10}$.}
\label{fig:parameter_effects}
\end{figure}

We show in Figure~\ref{fig:parameter_effects} that our recursive algorithm
behaves predictably. Empty bars in these plots indicate that the algorithm
failed to terminate for the corresponding pair of step size and $\theta_{max}$.
This behavior is not a limitation but is actually ideal, as these particular
parameter pairs produce biologically-implausible paths.  We see that the number
of turning angle sequences (i.e. the number of elements in $\mathit{\Sigma}$) increases
exponentially as step size goes down or $\theta_{max}$ increases.  The
estimated transition probabilities to edge and corner neighbors increases as
step size increases. Intuitively, this is a consequence of a step being able to
hop into neighboring edge or corner voxels from deeper in the source voxel
without first landing in one of the 6 face-sharing neighbors -- resulting in
larger potential starting cuboids ($\volume$) for those turning angle
sequences.

\newpage

\end{document}